\documentclass[11pt,a4paper]{article}
\usepackage{jcappub}

\usepackage{comment}
\usepackage[utf8]{inputenc} 
\usepackage{amsfonts,amssymb, mathrsfs, amsmath, esint,bm,enumitem}
\allowdisplaybreaks

\usepackage{subcaption}
\usepackage{ragged2e}
\DeclareCaptionJustification{justified}{\justifying}
\usepackage{diagbox}
\usepackage{latexsym}
\usepackage{graphicx}
\usepackage[dvipsnames]{xcolor}
\usepackage{booktabs,multirow}
\usepackage{titlesec} 
\usepackage{physics}
\usepackage[normalem]{ulem}

\usepackage{tikz}
\usetikzlibrary{positioning,calc}

\usepackage{hyperref}
\hypersetup{colorlinks, citecolor=ForestGreen, linkcolor=BrickRed, urlcolor=RedViolet}

\usepackage[capitalize,noabbrev]{cleveref}
\crefname{equation}{Eq.}{Eqs.}
\usepackage{float}
\usepackage{placeins}

\newcommand{\fede}{f_{\text{EDE}}}

\newcommand{\Neff}{N_\text{eff}}

\newcommand{\na}{$-$}

\newcommand{\corr}{R(H_0,\taureio)}

\newcommand{\baseline}{$\mathcal{B}$}
\newcommand{\baselinewoee}{$\mathcal{B}_\text{\sout{EE}}$}
\newcommand{\shoes}{$\mathbf{ H_0}$}
\newcommand{\desi}{\textbf{BAO}$_\textbf{DESI}$}
\newcommand{\bao}{$\mathbf{}$\textbf{BAO}}

\newcommand{\PTT}{Planck lowl TT}
\newcommand{\PEE}{Planck lowl EE}
\newcommand{\PTTTEEE}{Planck highl TTTEEE }
\newcommand{\act}{ACT/Planck Lensing}
\newcommand{\sixdf}{BAO 6dF}
\newcommand{\mgs}{BAO SDSS DR7 MGS}
\newcommand{\consensus}{BAO SDSS DR12}
\newcommand{\BAOdesi}{BAO DESI}
\newcommand{\pantheon}{SN Pantheon+}
\newcommand{\pantheonshoes}{SN Pantheon+SH0ES}

\newcommand{\DNeff}{\Delta N_\text{eff}}
\newcommand{\taureio}{\tau_{\rm reio}}
\newcommand{\omegacdm}{\omega_{\rm cdm}}
\newcommand{\Omegacdm}{\Omega_{\rm cdm}}

\newcommand{\beq}{\begin{equation}}
\newcommand{\eeq}{\end{equation}}

\title{Reionization and the Hubble Constant: Correlations in the Cosmic Microwave Background}

\author[a]{Itamar J.~Allali,}
\author[a]{Praniti Singh,}
\author[a,b]{JiJi Fan,}
\author[a]{Lingfeng Li}

\affiliation[a]{Department of Physics, Brown University, Providence, RI 02912, USA}
\affiliation[b]{Brown Theoretical Physics Center, Brown University, Providence, RI 02912, USA}
 
\emailAdd{itamar\_allali@brown.edu}
\emailAdd{praniti\_singh@brown.edu}
\emailAdd{jiji\_fan@brown.edu}
\emailAdd{lingfeng\_li@brown.edu}

\abstract{Recently, the James Webb Space Telescope (JWST) has found early galaxies producing photons from more efficient ionization than previously assumed. This may suggest a reionization process with a larger reionization optical depth, $\taureio$, in some mild disagreement with that inferred from measurements of cosmic microwave background (CMB). Intriguingly, the CMB would prefer larger values of $\taureio$, more consistent with the recent JWST hint, if the large-scale measurements (i.e. $\ell <30$) of E-mode polarization are removed. In addition,  $\taureio$ has an indirect correlation with today's Hubble constant $H_0$ in $\Lambda$CDM. Motivated by these interesting observations, we investigate and reveal the underlying mechanism for this correlation, using the CMB dataset without the low-$\ell$ polarization data as a proxy for a potential cosmology with a larger $\taureio$. We further explore how this correlation may impact the Hubble tension between early and late universe measurements of $H_0$, in $\Lambda$CDM as well as two proposals to alleviate the Hubble tension: the dark radiation (DR) and early dark energy (EDE) models. We find that the Hubble tension gets further reduced mildly for almost all cases due to the larger $\taureio$ and its positive correlation with $H_0$, with either the Baryon
Acoustic Oscillations (BAO) data before those from the Dark Energy Spectroscopic Instrument (DESI) or the DESI data. 
}

\begin{document}
\maketitle
\vspace{1em}\noindent

\section{Introduction}
\label{sec:intro}

A detailed understanding of the epoch of reionization and the ``cosmic dawn" is the focus of ongoing and future pursuits, especially the James Webb Space Telescope (JWST)~\cite{2022A&A...661A..80J,2023PASP..135d8003W}. This time period in the expansion history of the universe marks the ending of the ``dark ages," when the universe was filled with a mostly neutral gas, and the beginning of star and galaxy formation with the reionization of baryonic matter. Astrophysical observatories such as JWST can study this epoch by observing the earliest stars and galaxies, providing insight into the timing and duration of the reionization of the universe. In addition, the cosmic microwave background (CMB), the remnant light from the hot ionized plasma in the early universe and the most precise cosmological probe, provides a test of reionization via the measurement of 
$\taureio$, the optical depth of the reionized universe through which the CMB travels to our observatories. 

The direct effects of $\taureio$ on the CMB observables are quite straightforward. Long after the CMB is released during the epoch of recombination, the universe undergoes reionization beginning at a redshift $z < 30$ (even as late as $z<10$, see \cite{Planck:2018vyg}); CMB photons can be (Thomson) scattered in this reionized universe, slightly scrambling the power spectrum that was encoded at the surface of last scattering. This shows up as an overall suppression of the CMB multipole spectra by a factor of $e^{-2\taureio}$ primarily for multipoles $\ell \gtrsim 100$. This effect is almost degenerate with a shift in the overall amplitude of the primordial power spectrum $A_s$: the effective amplitude measured in the CMB anisotropies is $A_s e^{-2\taureio}$, and thus an increase in either $A_s$ or $\taureio$ can be compensated with an increase in the other. However, the large-scale modes outside the horizon at reionization are less suppressed by $\taureio$.
Modes that cross the horizon during reionization and later ($\ell \lesssim 100$) are subject to less Thomson scattering and thus the power spectrum is less suppressed at these scales.
In addition, the Thomson scattering in the late universe has an additional effect on the polarization of the CMB photons, enhancing the E-mode spectra on the largest scales ($\ell \lesssim 10$). These effects break the degeneracy between $A_s$ and $\taureio$ and can serve as an anchor to help determine $\taureio$ more precisely.

The astrophysical approach, on the other hand, serves as an indirect yet independent way of measuring $\taureio$~\cite{Robertson:2015uda,McQuinn:2015icp}. The photon rescattering effect is not directly observed in this case, in contrast to the CMB measurement. Instead, the overall $\taureio$ is inferred from the distributions of luminous galaxies, the production rate of ionizing photons in these galaxies, and the chance of their escaping their host galaxies. 

The availability of independent means of studying reionization presents an opportunity for confirming our understanding of this epoch or conversely gaining new insight should independent means of measurement disagree. In contrast to the measurement of $\taureio = 0.0544\pm 0.0073$ from the \textit{Planck} satellite's observations of the CMB \cite{Planck:2018vyg}, recent work suggests a higher $\taureio$ based on JWST data, which could be as high as $\taureio\gtrsim$ 0.07 (see~\cite{Munoz:2024fas} for various approaches resulting in values of $\taureio$ in this regime). In particular, the increase of $\taureio$ is mainly driven by a higher ionizing efficiency inferred from JWST observations~\cite{2024MNRAS.527.6139S,2024MNRAS.533.1111E}, associated with a higher photon production rate at early times. More observed star-forming galaxies at high $z$ also have a subdominant effect in increasing $\taureio$~\cite{2023ApJ...946L..13F,2024ApJ...969L...2F}. Furthermore, several recent surveys in the low-$z$ region~\cite{2022MNRAS.517.5104C} indicate a higher escape fraction and imply an even higher $\taureio \sim 0.096$ when the result is extrapolated to a higher $z$.

The studies based on JWST data could be subject to substantial uncertainties from observational bias in galaxy surveys, parameter extrapolation, and baryon physics modeling. See also later discussions~\cite{Mukherjee:2024cfq,Paoletti:2024lji,Zhu:2024xrt,Cain:2024fbi}. 
On the other hand,  the precise determination of $\taureio$ from CMB relies on the large-scale polarization data, which has the largest uncertainties among CMB measurements. Interestingly, Ref.~\cite{Giare:2023ejv} showed that excluding the large-scale CMB data generically predicts a larger value of $\taureio$. This is more consistent with the value inferred from JWST observations reviewed above. 

As these new observations stand to challenge our current picture of reionization, it is important to understand how these measurements may impact other aspects of the standard $\Lambda$CDM model of cosmology. In particular, as we will discuss in \cref{sec:physics}, the present expansion rate of the universe $H_0$ has an observed correlation with the value of $\taureio$ when inferred using CMB observations. Given the long-standing tension in the determination of $H_0$ from CMB observations versus more direct observations of distance ladder methods \cite{Riess:2021jrx,Breuval:2024lsv,Scolnic:2023mrv,Riess:2024vfa,H0LiCOW:2019pvv,Freedman:2024eph,Freedman:2023jcz,Freedman:2021ahq}, the precise value of $\taureio$ may be of interest. The correlation of these two parameters suggests that an increase in the inferred value of $\taureio$ may alleviate the ``Hubble tension."

In this work, we aim to explore the correlation of $\taureio$ and $H_0$ and to assess the degree to which this correlation may impact the tension in different measurements of $H_0$. In \cref{sec:methods}, we discuss the models that we will study in this work and outline the datasets we consider. Then, in \cref{sec:results_all} we discuss our results. \cref{sec:physics} first covers the physical mechanisms leading to the correlation between $H_0$ and $\taureio$ when inferred from CMB observations. Then, \cref{sec:results} gives a detailed discussion of inferred parameters when fitting various models with and without large-scale polarization data from the CMB in order to assess how shifts in $\taureio$ can impact shifts in $H_0$. Finally, in \cref{sec:conclusion}, we provide a concluding discussion. 

\section{Cosmological Models and Methods}
\label{sec:methods}

\subsection{Cosmological Models}
\label{sec:models}

Throughout this work, we will focus on the following cosmological models.\\

\noindent{\bf $\Lambda$CDM}\\

The $\Lambda$CDM model describes the universe as being composed of radiation, ordinary (baryonic) matter, and cold dark matter (CDM). Additionally, it includes a dark energy component, parameterized by the cosmological constant $\Lambda$, which drives the observed accelerated expansion of the universe. In this work we take the universe to be spatially flat.
 
We parametrize this framework with six major parameters: the Hubble rate of expansion today (Hubble constant) $H_0=100 \, h$ km/s/Mpc, the baryon density $\omega_b=\Omega_b h^2$ (where $\Omega_i\equiv \rho_i/\rho_{c}$ is the abundance for the $i^\text{th}$ species, $\rho_i$ its density, and $\rho_{c}$ the critical energy density for a flat universe today), the cold dark matter density $\omegacdm=\Omegacdm h^2$,  the scalar spectral index $n_s$, the amplitude of primordial scalar fluctuations $A_s$, and the reionization optical depth $\taureio$. Additionally, in our work, we model the Standard Model (SM) neutrinos as a combination of one massive species with a mass of 0.06 eV and two massless species. \\

\noindent{\bf Dark Radiation}\\

We also consider models of dark radiation (DR), 
additional light degrees of freedom  which remain ultra-relativistic until after the epoch of recombination  (for a review, see \cite{Archidiacono:2013fha}). Thus, in addition to photons and SM neutrinos, the DR component contributes to the radiation density $\rho_r$. The total relativistic degrees of freedom can be accounted for by the parameter $N_{\rm eff}$, related to $\rho_r$ as
\beq
\rho_r = \left[1+\frac{7}{8}\left(\frac{4}{11}\right)^{4 / 3} N_{\mathrm{eff}}\right] \rho_\gamma~,
\label{eq: Neff}
\eeq
where $\rho_\gamma$ is the energy density of photons fixed by the CMB temperature. The expression above is defined such that for a single fully thermalized neutrino species, $\Neff = 1$. However, when finite-temperature effects on photons and the non-instantaneous nature of neutrino decoupling are taken into account, the SM predicts $\Neff=3.044$ \cite{Akita:2020szl,Froustey:2020mcq,Bennett:2020zkv}. Thus, in the DR model, the contribution of additional relativistic species is parameterized using $\DNeff$, defined as:
\begin{equation}
\DNeff \equiv \frac{8}{7}\bigg(\frac{11}{4}\bigg)^{4/3}\frac{\rho_{\rm DR}}{\rho_\gamma}~,
\end{equation}
where $\rho_{\rm DR}$ is the energy density of DR. 

DR models can help resolve the Hubble tension by leading to a higher $H_0$ value inferred from CMB data. This is due to the DR's contribution to $\rho_r$ which reduces the size of the sound horizon and thus requires a higher $H_0$ to preserve its angular size (see e.g. \cite{DiValentino:2021izs,Schoneberg:2021qvd}).

Phenomenologically, the $\DNeff$ parameterization effectively encompasses a broad spectrum of particle physics models. However, the specific microphysical properties of DR can imprint distinct signatures on cosmological observables. In this work, we investigate two different DR scenarios:
\begin{enumerate}
    \item \textbf{Free-Streaming Dark Radiation (FSDR):} \\
    In this model, the relativistic species that make up the DR component either do not interact at all with themselves or other species, or have feeble interactions characterized by an interaction rate $\Gamma\ll H$. As a result, these species decouple from the primordial plasma early in the Universe's history, propagating freely similar to SM neutrinos. The free-streaming property of DR is well motivated by many particle physics models.\footnote{
    As there are many scenarios resulting in dark, light, non-interacting degrees of freedom, we direct the reader to reviews such as \cite{DiValentino:2021izs} for a comprehensive list of candidates.}

    \item \textbf{Self-Interacting Dark Radiation (SIDR):} \\
    Here, the DR species exhibit strong self-interactions such that their interaction rate $\Gamma \gg H$. This also captures the scenario where $\Gamma$ is large until after recombination.  These interactions inhibit free-streaming, resulting in a suppression of anisotropic stress and viscosity. In this limit, DR behaves as a perfect relativistic fluid, characterized by an equation of state $w = 1/3$. Such self-interaction may stem from various microscopic models such as non-abelian gauge bosons, an interacting light dark sector, or non-standard neutrino interactions with the dark sector (see further discussions in~\cite{Jeong:2013eza,Buen-Abad:2015ova,Buen-Abad:2017gxg,Brust:2017nmv,Blinov:2020hmc,Brinckmann:2022ajr,Buen-Abad:2024tlb}).
\end{enumerate}

\vspace{3mm}

\noindent{\bf Early Dark Energy}\\

Another model of interest is the Early Dark Energy (EDE) model~\cite{Karwal:2016vyq, Poulin:2018cxd}, which introduces a new component that behaves as dark energy in the early universe and subsequently decays around the time of recombination, effectively injecting energy into the universe at this epoch. This model is typically parameterized by a scalar field, $\phi$, with a potential given by $V(\theta) = m^2 f^2[1 - \cos(\theta)]^n$, where $m$ is the mass of the scalar field, $f$ is its decay constant, $\theta \equiv \phi/f$ and $n \geq 1$ is an integer. Initially, the scalar field is held at a fixed value, $\theta_i$, and behaves as a perfect fluid with an equation of state $w = -1$, keeping its background energy density nearly constant. When the Hubble parameter drops to about $H^2 \sim \partial_\theta^2 V(\theta)/f^2$, the field starts oscillating around the minimum of the potential. Once the oscillations begin around redshift $z_c$, the EDE component behaves as a fluid with an equation of state $w_n = (n-1)/(n+1)$. These oscillations cause the energy density of the field to dilute faster than that of radiation for $n > 2$.

 At the background level around recombination, the EDE model is analogous to the DR models: they both inject energy and increase the Hubble expansion parameter, $H(z)$, at that time. The resulting reduction in the sound horizon, $r_s$, is compensated by decreasing $D_A$, leading to a higher inferred value of $H_0$. The rapid dilution of EDE ensures that its effects are confined to the epoch around recombination, leaving the physics of the late-time universe unaffected.

We parameterize this model using the critical scale factor $a_c=1/(z_c+1)$ and the fraction contributed by EDE to the total energy density $\fede(z_c)$ at the critical redshift $z_c$. For our analysis, we fix $n=3$ and $\theta_i=2.72$, which are chosen based on the preference of the combined dataset CMB+SH0ES+BAO in~\cite{Smith_2020}.

\subsection{Data Analysis Methods}
\label{sec:data}
We use \texttt{CLASS}~\cite{Diego_Blas_2011,lesgourgues2011cosmiclinearanisotropysolving} to solve for the cosmological evolution in $\Lambda$CDM and DR scenarios, and \texttt{AxiCLASS}~\cite{Smith:2019ihp,Poulin:2018dzj} in the EDE case. We employ \texttt{Cobaya}~\cite{2019ascl.soft10019T,Torrado:2020dgo} to perform a Bayesian analysis by generating Markov Chain Monte Carlo (MCMC) samples. Posteriors and plots are obtained using \texttt{GetDist}~\cite{lewis2019getdistpythonpackageanalysing}.

To better discuss the impact on $\taureio$~from various data, we consider several combinations of datasets. First, we consider the following to be our ``baseline" dataset, against which we make most comparisons:
\begin{itemize}
    \item \baseline: CMB data from the \textit{Planck} satellite, including the public release 3 (PR3) likelihoods for the TT and EE power spectra at $\ell <30$~\cite{Planck:2019nip}, and the
    \texttt{CamSpec} likelihood~\cite{Efstathiou_2021} 
    based on
    the 2020 Planck PR4 data release~\cite{Planck:2020olo} by~\cite{Rosenberg_2022} for $\ell >30$ TT, TE, EE data. For CMB lensing, we use the Atacama Cosmology Telescope Data Release 6 (ACT DR6) lensing likelihood~\cite{ACT:2023dou,ACT:2023kun} combined with the \textit{Planck} PR4 lensing likelihood~\cite{Carron:2022eyg}. We also add the Pantheon+ Type Ia supernovae catalog~\cite{Scolnic:2021amr}.
\end{itemize}
As briefly mentioned in the introduction, the CMB constraints on $\taureio$ largely come from the EE spectrum at the largest scales. Thus, to isolate the impact from the low-$\ell$ EE spectrum, we introduce the modified baseline dataset:
\begin{itemize}
    \item \baselinewoee: The same as \baseline~but with the EE likelihood for $
\ell<30$ removed.
\end{itemize}
In addition to CMB and Type Ia supernovae data, we also include measurements of Baryon Acoustic Oscillations (BAO) as an important probe of the cosmological history. In the baseline dataset \baseline, the BAO data is not specified as the field is under active progress. The differences between BAO datasets are not yet fully understood. In order to ensure our discussion on $\taureio$ is not vulnerable to BAO details, different BAO datasets will be combined with the baseline datasets. We consider the previously accepted set of BAO measurements:
\begin{itemize}
     \item \bao: 
     BAO measurements from the 6dF Galaxy Survey (6dFGS) at $z = 0.106$~\cite{Beutler_2011}, the Sloan Digital Sky Survey (SDSS) Data Release 7 (DR7) for the main galaxy sample (MGS) at $z=0.15$~\cite{Ross:2014qpa}, and the CMASS and LOWZ samples of the SDSS BOSS DR12  at redshifts $z= 0.38$, $0.51$, and $0.61$~\cite{BOSS:2016wmc}.
\end{itemize}
Alternatively, we will consider the more recent BAO measurements from the Dark Energy Spectroscopic Instrument (DESI). In particular, the measurements from DESI have shown some preference for larger $H_0$ and some important impact on resolutions to the Hubble tension (see e.g. \cite{DESI:2024mwx,Allali:2024cji,Qu:2024lpx,Wang:2024dka,Seto:2024cgo,Lynch:2024hzh,Jiang:2024xnu,Chatrchyan:2024xjj,Li:2025rjr}).
\begin{itemize}
   \item \desi:  BAO measurements from DESI 2024 for effective redshifts $z = 0.3$, $0.51$, $0.71$,
 $0.93$, $1.32$, $1.49$, $2.33$ \cite{DESI:2024mwx}.
\end{itemize}
In the following sections, we will explore the impact of these choices of data on the inferences of $\taureio$ and $H_0$.


\section{Results}
\label{sec:results_all}
The results of our analysis are discussed below. We begin first by exploring in detail the correlation of $\taureio$ and $H_0$ in the $\Lambda$CDM model. Following that, we present detailed statistics from our analyses of each model discussed, fitting to several datasets. We will make comments on the inferences of $\taureio$ and $H_0$, and its impact on the Hubble tension.

\subsection{$\taureio$--$H_0$ Correlation Within $\Lambda$CDM}
\label{sec:physics}
\begin{figure}
    \centering
    \includegraphics[width=0.7\linewidth]{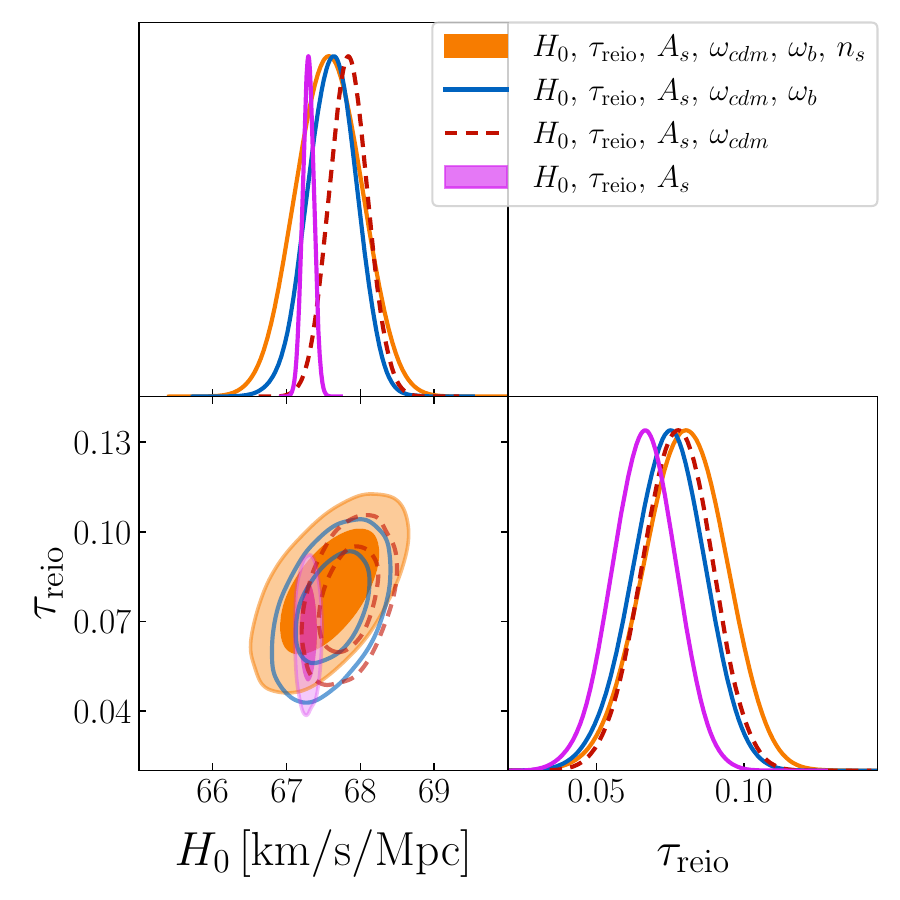}
    \caption{The one- and two-dimensional posterior distributions of $H_0$ and $\tau_{\rm reio}$ in a series of fits to the \baselinewoee+\bao~dataset. In the purple contour, only three parameters, $H_0$, $\tau_{\rm reio}$, and $A_s$, are allowed to vary. Other cosmological parameters are fixed to \textit{Planck} PR3 best fit values \cite{Planck:2018vyg} ($\omegacdm=0.12011$, $\omega_b=0.022383$, $n_s=0.96605$). In this case, $H_0$ is tightly bounded by the $\theta_s$ observable. The fit is followed by the 4-, 5-, and 6-parameter fits by adding $\omegacdm$, $\omega_b$, and $n_s$ as free parameters, respectively. The positive correlation between $H_0$ and $\tau_{\rm reio}$ increases after the inclusion of each parameter, indicating that their correlations are introduced via other cosmological parameters. }
    \label{fig:correlation}
\end{figure}

In contrast to the direct effects of $\taureio$ on the CMB spectra, the relationship between $\taureio$ and other parameters is not as straightforward. In particular, we are interested in understanding the correlation between the Hubble constant $H_0$ and $\taureio$. These two parameters are indirectly correlated, relying on other parameters in $\Lambda$CDM. In \cref{fig:correlation}, we show a set of two-dimensional posterior distributions for $H_0$ and $\taureio$ in the $\Lambda$CDM model, fit to the \baselinewoee+\bao~dataset as defined in \cref{sec:data}, to visualize the correlations.

\cref{fig:correlation} shows in purple that the inferred values of $H_0$ and $\taureio$ are uncorrelated when only these two parameters, and the amplitude $A_s$, are left free to vary. This is partially due to the fact that $H_0$ is nearly fixed by the angular size of the sound horizon at recombination $\theta_s$, which corresponds to the location of the first peak in the CMB power spectrum, when the total matter density $\omega_{m}=\Omega_m h^2$ is fixed (see below). Thus, if we allow the cold dark matter density $\omegacdm$ to vary, $H_0$ becomes more variable, and the correlation with $\taureio$ appears. This is seen in \cref{fig:correlation} in the red dashed contours. We can quantify the correlation by computing the Pearson correlation coefficient $\corr$ (the ratio of the parameters' covariance to the product of their standard deviations). For the case of the 4-parameter fit, $\corr=0.34$. Adding the baryon density $\omega_b$ as a free parameter allows for an even greater variation of $H_0$ without significantly enhancing the correlation with $\corr=0.38$. Finally, adding the final free parameter of $\Lambda$CDM, $n_s$, also adds to the degree of correlation, giving the orange filled contours in \cref{fig:correlation} and $\corr=0.57$. Note that the data that results in these correlations excludes the low-$\ell$ EE CMB data, which allows for a greater range of $\taureio$ and aids in understanding the correlation. In fact, as we will explore in the remainder of this section, the effects driving this correlation can be understood via primarily the TT part of the CMB power spectrum at $\ell > 30$, and fitting to this data alone gives an even stronger correlation of $\corr=0.75$. In \cref{sec:results}, we explore other effects of the datasets further.

\begin{figure}
    \centering
    \includegraphics[width=0.53\linewidth]{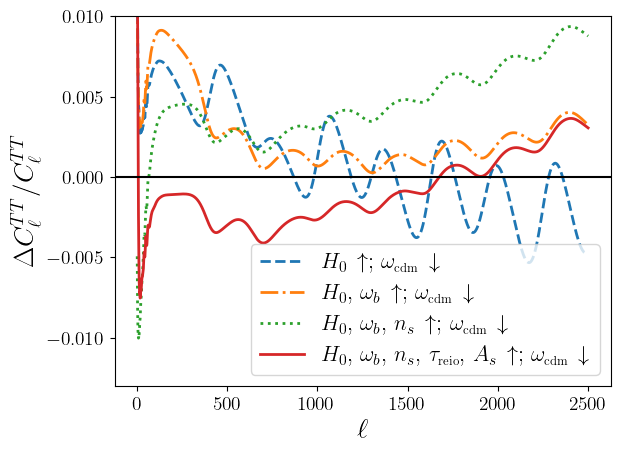}
    \includegraphics[width=0.42\linewidth]{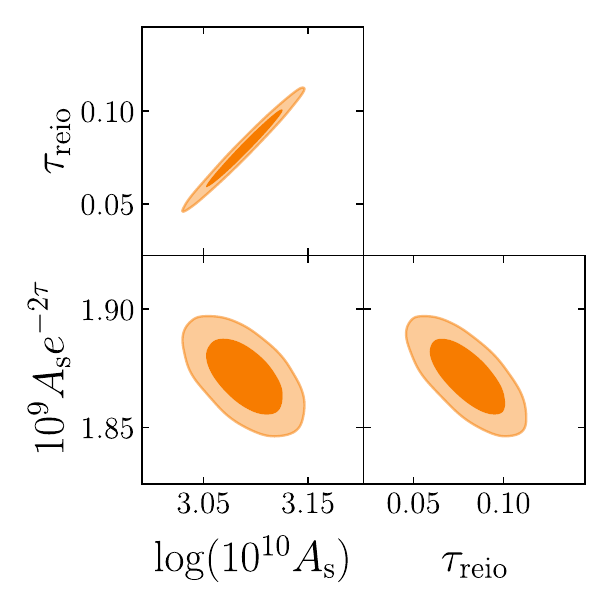}
    \caption{{\bf (Left)} The relative shift of the CMB TT power spectrum between two benchmark sets of parameters taken from the Monte Carlo sample shown in orange in \cref{fig:correlation} fit to the \baselinewoee+\bao~dataset. These two sets of parameters are chosen to highlight the shift in $H_0$, and both have a nearly minimal effective $\chi^2$ and thus represent cosmologies that fit the data reasonably well. The parameters are given in \cref{tab:parametersshift}. In the legends, $\uparrow$ indicates parameters increased with respect to the initial choice, while $\downarrow$ indicates a decrease.
    \\{\bf (Right)}Two-dimensional posterior distributions of $\{\taureio,A_s,A_se^{-2\taureio}\}$ from the 6-parameter fit to the \baselinewoee+\bao~dataset depicted in \cref{fig:correlation}. The tight correlation of $\taureio$ and $A_s$ necessitates their simultaneous shift, while their anticorrelation with the effective CMB amplitude $A_se^{-2\taureio}$ indicates that a simultaneous increase of $A_s$ and $\taureio$ results in a suppression of the CMB power spectrum, as in the red curve of the left panel.}
    \label{fig:DeltaCl}
\end{figure}

Let us now explore how the correlation between $H_0$  and $\taureio$ comes about from the variation of other parameters. First, in order to increase $H_0$ while keeping the angular size of the sound horizon $\theta_s$ the same, we must decrease $\omega_m$ in the flat $\Lambda$CDM model.
One can compute this scale directly from the values of $H_0 = 100 h \,\mbox{ km/s/Mpc}$ and $\omega_m = \Omega_m h^2$, given by the ratio of the sound horizon at recombination $r_s$ and the angular diameter distance to the surface of last scattering $D_A$
\beq
\theta_s = \frac{r_s}{D_A} = \frac{\int^\infty_{z_\text{rec}} c_s \, dz/H(z)}{\int^{z_\text{rec}}_0 c\, dz/H(z)}~,
\label{eqn: sound horizon}
\eeq
where $c_s$ is the speed of sound in the early universe (dependent on $\omega_b$), $z_\text{rec}$ is the redshift of recombination, and $H(z)$ is given by the Friedmann equation 
\beq H^2(z) = (100 \mbox{ km/s/Mpc})^2 [\omega_r (1+z)^4 +\omega_m (1+z)^3 + h^2-\omega_m-\omega_r]~.
\label{eq: Hubble eqn}
\eeq
Note that the radiation abundance $\omega_r=\Omega_rh^2$ is relatively fixed in $\Lambda$CDM by the CMB temperature and the number of neutrinos. An increase in $H_0$ for fixed $\omega_m$ keeps $r_s$ virtually unchanged, while decreasing $D_A$; thus, a decrease in $\omega_m$ is needed to increase $D_A$ and leave $\theta_s$ unchanged. 

Increasing $H_0$ and decreasing $\omega_m$ both induce opposite phase shifts (by shifting the primary peak location corresponding to $\theta_s$) in the CMB spectra, which nearly cancel. However, some major deformations of the spectrum remain. First, there is a leftover enhancement of the CMB spectrum in the vicinity of the first peak, around $\ell\sim 200$. This change can be seen in the blue dashed curve in the left panel of \cref{fig:DeltaCl}, depicting the residual change in the multipole coefficients $C_\ell$ of the TT power spectrum when increasing $H_0$ and decreasing $\omegacdm$ with respect to a reference choice of parameters (given in \cref{tab:parametersshift}). With a decrease in $\omega_m$, the time of matter-radiation equality is delayed, resulting in two reasons for an enhancement of the primary peak: (i) there is less time during matter-domination prior to recombination for modes crossing the horizon around that time to be damped; and (ii) the early integrated Sachs-Wolfe (EISW) effect is enhanced because the metric potentials have less time to settle to their constant values, further enhancing the peak (see \cite{Lesgourgues:2013qba} for more details).
Another deformation to the spectrum caused by the shift in $H_0$ and $\omegacdm$ is a residual phase shift due to the relative heights of even and odd peaks in the power spectrum. This can be compensated by shifting the baryon abundance $\omega_b$, as shown in the orange dot-dashed curve of the left panel in \cref{fig:DeltaCl}.

\begin{table}[h!]
    \centering
    \begin{tabular}{c|cccccc}
    \hline
    \hline
    Parameters  & $H_0 \,[\mathrm{km}/\mathrm{s}/\mathrm{Mpc}]$  & 
    $\omegacdm$ & $\omega_b$ & $n_s$ & $A_s$ & $\tau_{\rm reio}$  \\\hline
       Unmodified  &  67.4 & 0.1193 & 0.02218  & 0.9632 & 2.21$\times 10^{-9}$ & 0.083\\
       Modified  & 68.1 & 0.1179 & 0.02234 & 0.9672 & 2.25$\times 10^{-9}$ & 0.095\\
       \hline
       \hline
    \end{tabular}
    \caption{The parameters before and after the modifications corresponding to the curves in the left panel of \cref{fig:DeltaCl}; the power spectrum difference from shifting all of these parameters at once is given by the red curve.}
    \label{tab:parametersshift}
\end{table}

With the phase of the peaks mostly restored, we are still left with an increase in the CMB anisotropy power at $\ell < 500$ at around the percent level. This could be somewhat corrected by an increase in $\taureio$ which suppresses the power spectrum, explaining the very mild correlation in the four- and five-parameter fits of \cref{fig:correlation}. However, since the largest enhancement of power remains at smaller $\ell$, this introduces an effective red tilt to the power spectrum residuals (higher power at large scales/small $\ell$); thus $\taureio$, which suppresses power nearly democratically at most scales, cannot compensate directly. Instead, one can increase the spectral index $n_s$, which controls precisely the tilt of the power spectrum, introducing a relative blue tilt (shifting the already red tilt closer to scale invariance). This effect is shown in the green dotted curve of the left panel in \cref{fig:DeltaCl}, pushing down the TT residual for the largest scales (smallest $\ell$). Now, an increase in $\taureio$ is appropriate to suppress the power spectrum. In practice, the fit to data prefers both an increase in $\taureio$ and $A_s$, corresponding to the red curve in \cref{fig:DeltaCl} with residuals well below the percent level down to $\ell < 30$. The initial and final sets of parameters, resulting in the red residual curve, are given in \cref{tab:parametersshift}.

We want to comment more on the last step above in which the data prefer to raise both $\taureio$ and $A_s$ simultaneously, instead of shifting only one of these two nearly degenerate parameters. Examining the contours in the right panel of \cref{fig:DeltaCl}, from the same analysis as in \cref{fig:correlation}, we observe the correlations between $A_s$, $\taureio$, and the effective amplitude measured in the CMB $A_s e^{-2\taureio}$. First, one notices the very tight correlation between $\taureio$ and $A_s$. Since the effective amplitude of the CMB is measured relatively precisely, shifting $\taureio$ necessitates shifting $A_s$ to restore the measured value of $A_s e^{-2\taureio}$; thus, both parameters must shift in unison.  Then, after shifting $H_0$, $\omegacdm$, $\omega_b$, and $n_s$, one is left with an overall enhanced power spectrum with respect to the initial benchmark. \cref{fig:DeltaCl} shows that achieving this suppression is possible with an increase in $A_s$ and $\taureio$, as evidenced by their anticorrelation with the effective amplitude.

We can therefore appreciate that the correlation between $H_0$ and $\taureio$ within the flat $\Lambda$CDM model crucially involves the other parameters in the model. Having established the mechanism for this correlation, we wish to understand the impact that an increase in $\taureio$ could have on inferences of  $H_0$ and the Hubble tension, as explored in the remainder of this work.

 
\subsection{Data Impact on $\taureio$ and the $H_0$ Tension}
\label{sec:results}
We will focus on constraints for $\left\{H_0, \taureio\right\}$, as our aim is to examine the role of inferences of $\taureio$ in determining the value of $H_0$ and consequently the Hubble tension. Since the removal of large-scale polarization data results in an increase in the inferred value of $\taureio$, we will make a comparison between fitting to the \baseline~and the \baselinewoee~datasets to assess the impact of such an increase. In each case, we combine with either \bao~or \desi.

We present the complete results of our analyses in \cref{appendix} for all models and dataset combinations discussed above. In addition to the analyzing \baseline+ \bao\ and \baselinewoee+\bao\ datasets, we also extend our investigation to include the SH0ES measurement of $H_0$~\cite{Riess:2021jrx}, see \cref{appendix} for more details. For 
each analysis, we provide posterior distributions and 68\% C.L. intervals for all six cosmological parameters $\left\{ A_\mathrm{s},~n_{\mathrm{s}},~\taureio,~H_0, ~\omegacdm, ~\omega_b \right\}$, along with new physics parameters for both the DR and EDE models, in \cref{appendix}. Furthermore, we provide best-fit values of these parameters along with the corresponding $\chi^2$ values for the different likelihoods used in our analyses in \cref{appendix}.

\begin{figure}[]
    \centering
    \includegraphics[scale=0.6]{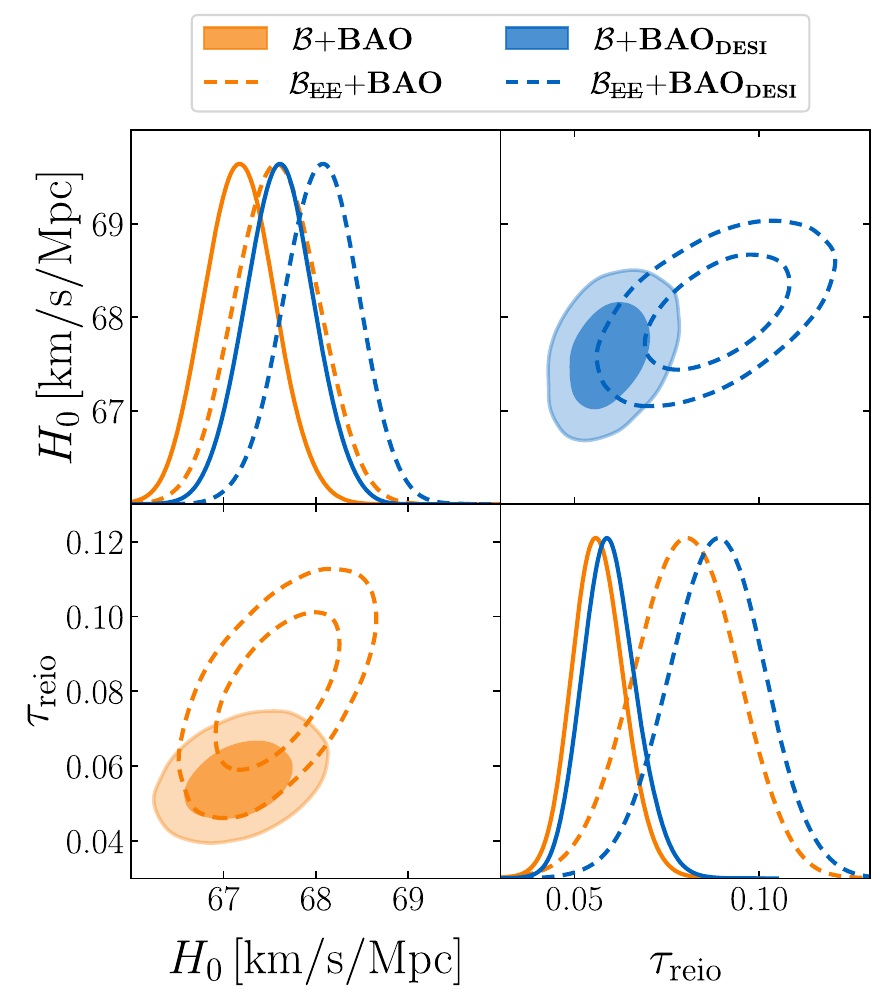}
\caption{Posterior distributions of $H_0$ and $\taureio$ in $\Lambda$CDM. The results from the \bao~dataset are shown in the lower-left part of the plot as orange contours. In the upper-right corner, results from the \desi~ dataset are shown in blue. Solid curves and filled contours represent fits to the \baseline~dataset, while dashed curves/contours represent the outcome from the \baselinewoee~dataset. The \desi~dataset tends to have higher $H_0$ and $\taureio$ central values than the \bao~dataset, with their correlations similar in both cases. Removing the low-$\ell$ EE data significantly shifts the central values of $\tau_{\rm reio}$ and enlarges the corresponding uncertainties. The central values and uncertainties of $H_0$ also increase, though less significantly. }
\label{fig:LCDM_quadrilateral}
\end{figure}

\subsubsection{$\Lambda$CDM}
\label{sec: result_lambdacdm}
\cref{fig:LCDM_quadrilateral} shows posteriors of  $\left\{H_0,\taureio\right\}$ in $\Lambda$CDM, and their allowed ranges at 68\% C.L. are listed in \cref{Table:LCDM}.
As discussed in the previous sections, for modes within the horizon at reionization, corresponding to $\ell\gtrsim100$, the power spectrum is suppressed by a factor $e^{-2\taureio}$ and the associated data only depends on the combination $A_{s}e^{-2\taureio}$. This degeneracy is broken by the low-$\ell$ data, in particular, the polarization measurements.  Thus, from \cref{fig:LCDM_quadrilateral}, we can see that $\taureio$ is more constrained in the presence of low-$\ell$ EE CMB spectra data (solid contours). Without the low-$\ell$ ($\ell < 30)$ EE CMB spectra, there could be more variation in $\taureio$ with a larger mean value, as well as a stronger correlation between $\taureio$ and $H_0$. This correlation is indirect and arises through an involved interplay with other cosmological parameters, as discussed in detail in \cref{sec:physics}. We also provide in \cref{Table:LCDM} the Pearson correlation coefficient between $H_0$ and $\taureio$ denoted as $\corr$.

\cref{Table:LCDM} also lists the Gaussian tension  when comparing the values of $H_0$ in our runs with the SH0ES measurement $(H_0)_{\text{SH0ES}} =73.04$ km/s/Mpc with its  standard deviation, $\sigma_{\text{SH0ES}}=1.04$ km/s/Mpc \cite{Riess:2021jrx}. The Gaussian tension for a given run is calculated as:
\beq
(\Delta H_0)_{\text{GT} }= \frac{\left|H_0-(H_0)_{\text{SH0ES}}\right|}{\sqrt{\sigma^2+\sigma_{\text{SH0ES}}^2}}~,
\label{eq:H0tension}
\eeq
where $H_0$ is the mean value and $\sigma$ the standard deviation from the MCMC analysis. One can see that for a given BAO dataset, without the low-$\ell$ polarization data, the Hubble tension is mildly alleviated, thanks to the larger $\taureio$ and its positive correlation with $H_0$.

It has already been shown that using the DESI BAO measurements tends to enhance the value of $H_0$ \cite{DESI:2024mwx,Allali:2024cji,Qu:2024lpx,Wang:2024dka,Seto:2024cgo,Lynch:2024hzh,Jiang:2024xnu,Chatrchyan:2024xjj,Li:2025rjr}. Here we also show our results for runs with \baseline+\desi, observing slightly higher preferred values for both $\taureio$ and $H_0$. Furthermore, upon removing the low-$\ell$ EE data, these parameters shift to even higher values, and the Hubble tension is reduced down to $4.5\sigma$ when fit to \baselinewoee+\desi. Moreover, for both \bao~and \desi~runs, we find that the correlation between $\taureio$ and $H_0$ remains nearly the same, with $\corr=0.58$ for \baselinewoee+\bao~and $\corr=0.53$ for \baselinewoee+\desi.

\begin{table}[]
\centering
\begin{tabular} { l |c c c c}
\hline
\hline
{\bf } &  {\bf \baseline +\bao} &  {\bf \baselinewoee +\bao} &  {\bf \baseline +\desi} & {\bf \baselinewoee +\desi}\\
\hline
$\tau_\mathrm{reio}$ & $0.0565^{+0.0066}_{-0.0074}$ & $0.080\pm 0.014            $ & $0.0596^{+0.0066}_{-0.0078}$ & $0.088\pm 0.013            $\\
{$H_0 \,[\mathrm{km}/\mathrm{s}/\mathrm{Mpc}]           $}                     & $67.18\pm 0.38             $ & $67.58\pm 0.44             $ & $67.60\pm 0.37             $ & $68.05\pm 0.40             $\\
$\corr$ & 0.34 & 0.58 & 0.34 & 0.53 \\
$H_0$ Tension& $5.3\sigma $ & $4.8\sigma $ & $4.9\sigma $ & $4.5\sigma $ \\
\hline
\hline
\end{tabular}

\caption{The posterior central values and corresponding 68\% C.L. intervals from the $\Lambda$CDM model, fitting to four different datasets. The Pearson correlation coefficient $\corr$ is given. All $H_0$ tension values are obtained from the Gaussian approximation in \cref{eq:H0tension}.}
\label{Table:LCDM}
\end{table}

\begin{table}[b]
\centering
\begin{tabular} {l | c c | c c}
\hline\hline
 & \multicolumn{2}{c|}{FSDR} & \multicolumn{2}{c}{SIDR}\\
{\bf } &  {\bf \baseline +\bao} &  {\bf \baselinewoee +\bao} &  {\bf \baseline +\bao} & {\bf \baselinewoee +\bao}\\
 
\hline
$\tau_\mathrm{reio}$ & $0.0566\pm 0.0071          $ & $0.078\pm 0.014            $ & $0.0570\pm 0.0071          $ & $0.081\pm 0.014            $\\

$H_0 \,[\mathrm{km}/\mathrm{s}/\mathrm{Mpc}]           $                     & $68.13^{+0.58}_{-0.92}     $ & $68.45^{+0.60}_{-0.91}     $ & $68.14^{+0.59}_{-0.94}     $ & $68.69^{+0.69}_{-1.1}      $\\
$\corr$ & 0.17 & 0.27 & 0.23 & 0.35 \\
$\Delta N_{\mbox{eff}}$ (95\% C.L.)   & $< 0.376                   $ & $< 0.370                   $ & $< 0.340                   $ & $< 0.380                   $\\
$H_0$ Tension& $4.1\sigma $ & $3.8\sigma $ & $4.1\sigma $ & $3.5\sigma $ \\
\hline\hline
\end{tabular}
\caption{The posterior mean values and corresponding 68\% C.L. intervals for SIDR and FSDR models for parameters $\left\{H_0, \taureio \right\}$, along with $H_0$ tension values. For the new physics parameter $\DNeff$, the 95\% C.L. upper limits are presented instead. The Pearson correlation coefficient $\corr$ is also given.}

\label{Table: DR model}
\end{table}

\subsubsection{Dark Radiation}
\label{sec: result_dr}
\label{subsec: DR}

\begin{figure}[]
    \centering
    \includegraphics[scale=0.6]{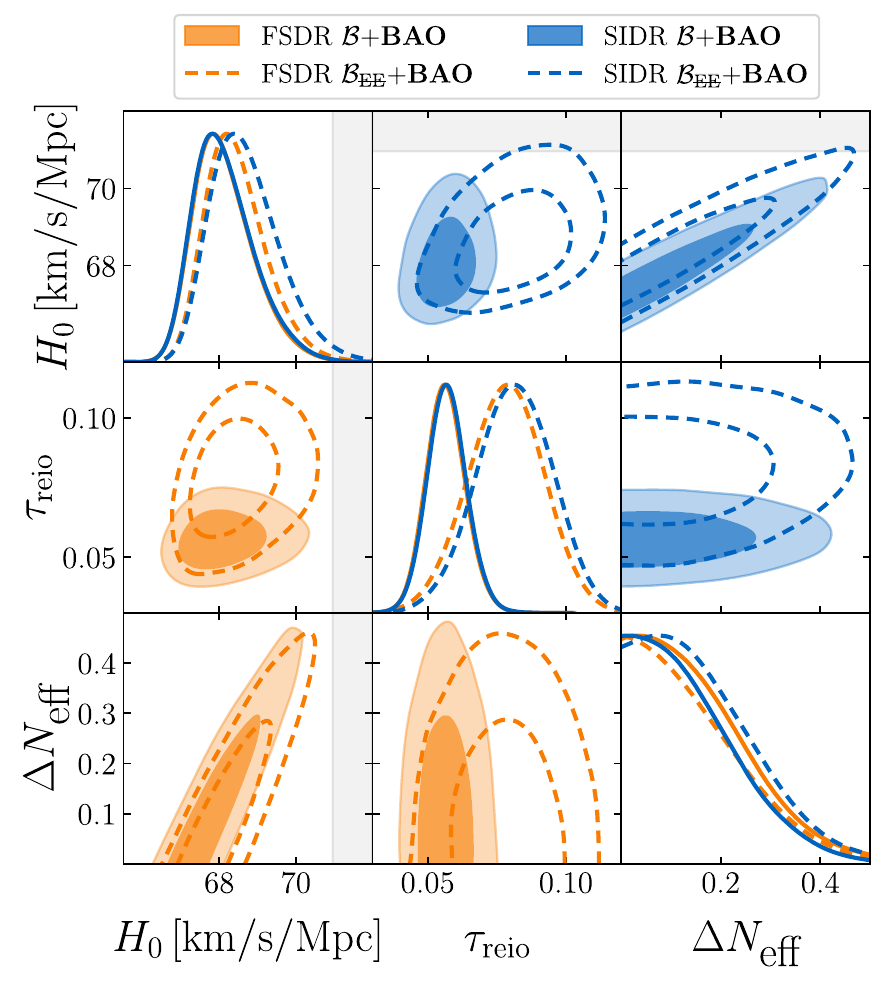}
    \caption{Posterior distributions of $H_0$, $\taureio$ and $\Delta N_{\rm eff}$ in DR models. The orange (blue) contours represent the free-streaming (self-interacting) scenario. Solid curves and filled contours represent fits to the \baseline~dataset, while dashed curves/contours represent the outcome from \baselinewoee~dataset. The gray shaded area shows the 2$\sigma$ range of $H_0$ from the SH0ES measurement. Similar to \cref{fig:LCDM_quadrilateral}, removing the low-$\ell$ EE data makes the central values and uncertainties of $\tau_{\rm reio}$ and $H_0$ larger. However, compared to the $\Lambda$CDM case, the correlation between $H_0$ and $\tau_{\rm reio}$ is weakened in both DR models.
    }
    \label{fig:DR_quadrilateral}
\end{figure}

An interesting feature of DR models is their potential to alleviate the  $H_0$ tension, as the presence of DR results in a higher inferred value of $H_0$ compared to $\Lambda$CDM.
The contribution of DR to $\omega_r$ leads to an increase in the expansion rate $H(z)$ through \cref{eq: Hubble eqn} at early times before recombination. Consequently, this increase in $H(z)$ reduces the sound horizon, $r_s$, at recombination. Thus, to maintain the angular size of the sound horizon, $\theta_s=r_s/D_A$, a higher value of $H_0$, which decreases $D_A$, is needed.

 Posteriors for the key parameters $\left\{H_0,\taureio,\DNeff \right \}$ for both the FSDR and SIDR models are shown in \cref{fig:DR_quadrilateral}, with 68\% C.L. constraints for $\left\{H_0, \taureio \right\}$ and 95\% C.L. upper bounds for $\DNeff$ presented in \cref{Table: DR model}. We provide the 95\% C.L. upper bounds for any parameter which does not generate a credible 68\% C.L. interval away from zero. From \cref{fig:DR_quadrilateral}, we can see the expected strong correlation between $H_0$ and $\DNeff$ for all four runs based on different combinations of datasets, due to the relatively fixed angular sound horizon scale $\theta_s$. For $\taureio$, both DR models exhibit a behavior similar to that observed in the $\Lambda$CDM model (see~\cref{sec: result_lambdacdm}): $\taureio$ is less constrained and tends to favor a higher value in the absence of low-$\ell$ EE CMB data. While $H_0$ and $\taureio$ are still correlated after removing the low-$\ell$ EE data (for similar reasons as discussed in \cref{sec:physics}), the strength of this correlation is now reduced, compared to those in $\Lambda$CDM model (see \cref{Table: DR model}). 
 The reduction in correlations can be understood as follows: the shift in $H_0$, $\DNeff$, $\omegacdm$, $\omega_b$, and $n_s$ that is induced in this scenario leaves the CMB power spectra overall suppressed, rather than mostly enhanced as in the case of $\Lambda$CDM (see \cref{sec:physics}), with the strongest suppression at the lowest $\ell \lesssim 100$. Thus, an increase in $A_s$ is warranted to restore the power spectrum. A further increase in $\taureio$ can be combined with the enhancement from $A_s$ so that the power is enhanced more at $\ell \lesssim 100$ than above, which nearly restores the power spectrum at all $\ell$.

Both DR models yield  similar constraints with the full CMB dataset. Yet the SIDR model exhibits a preference for slightly larger values of both $\taureio$ and $H_0$ in the absence of the low-$\ell$ CMB EE data compared to the FSDR model. Moreover, as expected, even with the low-$\ell$ EE measurements included, both models already favor slightly higher values of $H_0$ due to extra energy injection from additional relativistic species, i.e. $\DNeff>0$.  Removing the low-$\ell$ EE polarization data further amplifies this preference, thereby providing a moderately improved alleviation of the $H_0$ tension. This is also reflected in the reduction in the Gaussian tensions shown in the last line of \cref{Table: DR model}, once the low-$\ell$ EE data is removed.

The results for the DR models with \desi~runs are reported in~\cref{app: DR}. Similar to the $\Lambda$CDM results, we see slightly higher preferred values for both $H_0$ and $\taureio$. As shown previously in the literature \cite{Allali:2024cji}, SIDR has the potential to resolve the Hubble tension if the DESI results are confirmed in future releases, and thus this scenario is of interest. See \cref{app: DR} for more details.

\subsubsection{Early Dark Energy }
\label{sec: result_ede}

\begin{figure}[]
    \centering
    \includegraphics[scale=0.6]{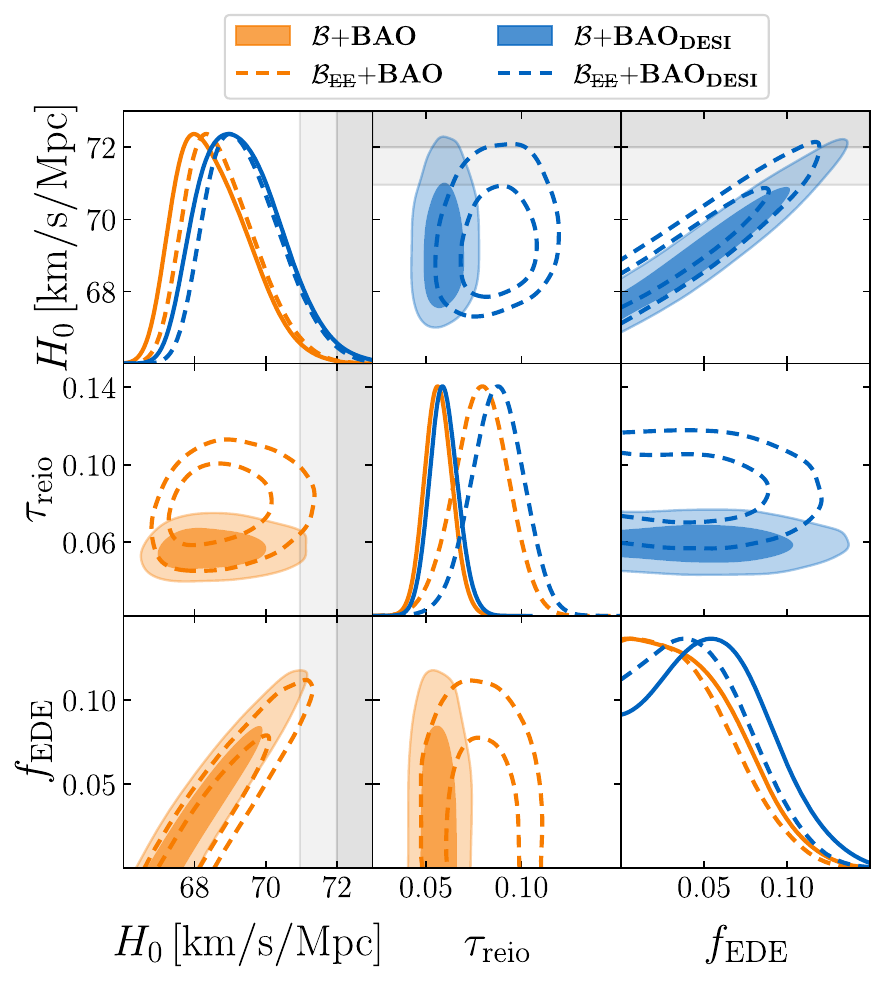}
    \caption{Posterior distributions of $H_0$, $\taureio$ and $f_{\rm EDE}$ in EDE models. The orange (blue) contours represent fits to the \bao~(\desi) dataset. Solid curves and filled contours represent fits to the \baseline~dataset, while dashed curves/contours represent the outcome from \baselinewoee~dataset.  The gray shaded area shows the $1\sigma$ (dark gray) and $2\sigma$ (light gray) range of $H_0$ from the SH0ES measurement.  Similar to \cref{fig:LCDM_quadrilateral}, removing the low-$\ell$ EE data makes the central value and uncertainties of $\taureio$ larger, while not for $H_0$. However, compared to the $\Lambda$CDM case, the correlation between $H_0$ and $\taureio$ is significantly weakened. 
    }
    \label{fig:EDE_quadritaleral}
\end{figure}

Here, posteriors of the parameters $\left\{H_0,\taureio, \fede \right \}$ are shown in \cref{fig:EDE_quadritaleral}, with their 68\% C.L. constraints presented in \cref{Table: EDE} for both the \baseline~and \baselinewoee~datasets combined with both \bao~and \desi. We provide the 95\% C.L. upper bounds for any parameter which does not generate a credible 68\% C.L. interval away from zero.
First, we see a strong correlation between $\fede$ and $H_0$ as expected. Since the EDE model is known to alleviate the $H_0$ tension, we observe a significant relaxation in the $H_0$ tension both with and without the low-$\ell$ EE polarization data. In particular, with the DESI BAO data, the $H_0$ tension is relaxed further compared to the results with earlier BAO data. For each choice of the BAO data, excluding the low-$\ell$ polarization data does not vary the level of Hubble tension significantly.

As observed in previous models, we find that $\taureio$ is less constrained, with higher values preferred when low-$\ell$ EE CMB spectral data are excluded. However, unlike previous models, we observe that even without the low-$\ell$ EE polarization data, the positive correlation between $\taureio$ and $H_0$ remains quite insignificant (see \cref{Table: EDE}). This is likely due to the fact that the increase in $H_0$ in the EDE model is mainly driven by $f_{\rm EDE}$. Thus, an increase in $\taureio$ could not help further increase $H_0$, which is consistent with the roughly constant Gaussian tension for a given BAO dataset, as reported in \cref{Table: EDE}.

\begin{table}[]
\centering
\begin{tabular} { l | c c c c}
\hline\hline
{\bf } &  {\baseline +\bao} &  {\baselinewoee +\bao} &  {\baseline +\desi} &  {\baselinewoee +\desi}\\

\hline
{$\taureio$} & $0.0566\pm 0.0073          $ & $0.079\pm 0.014            $ & $0.0594^{+0.0069}_{-0.0077}$ & $0.087\pm 0.014            $\\
{$H_0 \,[\mathrm{km}/\mathrm{s}/\mathrm{Mpc}]           $} & $68.52^{+0.77}_{-1.2}      $ & $68.78^{+0.74}_{-1.2}      $ & $69.32^{+0.92}_{-1.3}      $ & $69.43^{+0.80}_{-1.2}      $\\
$\corr$ & 0.11 & 0.22 & 0.10 & 0.18 \\
$\fede$         & $< 0.0977                  $ & $< 0.0924                   $ & $0.056^{+0.028}_{-0.041}   $ & $0.048^{+0.016}_{-0.045}   $\\
$H_0$ Tension& $3.5\sigma $ & $3.3\sigma $ & $2.7\sigma $ & $2.8\sigma $ \\
\hline\hline
\end{tabular}
\caption{The posterior mean values and corresponding 68\% C.L. intervals for the EDE model for parameters $\left\{H_0, \taureio, \fede \right\}$ along with its $H_0$ tension values. For $\fede$, the 95\% C.L. upper bounds are provided only when a credible 68\% interval away from zero is not obtained. The Pearson correlation coefficient $\corr$ is also given.}
\label{Table: EDE}
\end{table}

\section{Conclusion}
\label{sec:conclusion}

In this work, we have explored the correlation between the cosmological parameters $\taureio$ and $H_0$ in the context of $\Lambda$CDM cosmology and beyond. We have drawn on three motivations to guide this endeavor: 

{\bf (1)} Since its launch, JWST has been providing an unprecedented sensitivity to directly probe the properties of the first galaxies. Early observations reveal that faint, early galaxies were efficient producers of ionizing photons, with likely non-negligible escape fractions, suggesting an excess in the ionizing-photon budget during the epoch of reionization. These observations hint at a larger value of $\taureio$ than that inferred from the CMB.

{\bf (2)} The CMB constraints on $\taureio$ mainly come from E-mode polarization data at large scales, i.e., $\ell \lesssim 30$, by breaking the measurement degeneracy between $A_s$ and $\taureio$. Previously, ~\cite{Giare:2023ejv} showed that the CMB tends to prefer larger values of $\taureio$ when large-scale measurements are excluded.

{\bf (3)} There is an indirect correlation in the CMB data between inferences of $\taureio$ and $H_0$. Therefore, since a shift in $\taureio$ arising from either motivation (1) or (2) above could impact the inferred value of $H_0$ according to (3), we have investigated the nature of this correlation and have analyzed to what extent it can be informative for the Hubble Tension.

First, as expected, the absence of low-$\ell$ E-mode polarization data allows for greater variations in $\taureio$ across all three models we have studied. We observe a shift of approximately $1.4-2\sigma$ in each case. 
 In the $\Lambda$CDM model, this increased variation in $\taureio$ allows for a stronger positive correlation between $\taureio$ and $H_0$.  This correlation still exists but is weaker in the DR models compared to $\Lambda$CDM. In the EDE model, however, the correlation almost disappears.
 
For $\Lambda$CDM, as a result of the preference for larger values of $\taureio$ without large scale E-mode polarization data, we find that the Hubble tension gets reduced by a slight shift from $5.3 \sigma$ to $4.8 \sigma$ when combined with the \bao~dataset and $4.5\sigma$ with the \desi~dataset. While the DR and EDE models already mitigate the Hubble tension when the low-$\ell$ E-mode polarization data is included, we find that the tension is further reduced when excluding this dataset for almost all cases (except for the EDE model fit with the DESI BAO data).

To further contextualize our findings, a few concluding comments are in order:
\begin{itemize}
\item As emphasized in \cite{Giare:2023ejv}, the EE signal constraining the precise value of $\taureio$ is at the scales where the limit on the precision is set by cosmic variance, thus making results highly sensitive to even small unknown systematics. However, we do not claim to have any evidence against the low-$\ell$ EE dataset. Rather, we use the CMB dataset without the low-$\ell$ EE data as a proxy to study a potential cosmology with a larger $\taureio$. 

\item There are some other potential concerns on the large-scale polarization data~\cite{Giare:2023ejv}. Galactic foregrounds impact polarization anisotropies and can introduce potential errors if poorly understood. In addition, anomalies persist in temperature and polarization anisotropies at large scales including features in the TT and TE spectra at low multipoles ($\ell \lesssim 10$). 

\item Since the CMB EE and TE spectra at low multipoles could be significantly influenced by $\taureio$, precise measurements of these spectra can provide a clearer picture on $\taureio$ and reionization physics. Future CMB missions such as LiteBIRD~\cite{LiteBIRD:2020khw}, ECHO~\cite{Sen:2022usj,Adak:2021lbu} and PICO~\cite{PICO:2019thw} will be able to measure polarization across large scales, providing even tighter constraints on $\taureio$. In addition, ground-based experiments such as Simons Observatory (SO)~\cite{SimonsObservatory:2018koc}, POLARBEAR~\cite{POLARBEAR:2016wwl}, and BICEP3-Keck array~\cite{Bicep} will provide complementary observations at smaller scales leading to a more complete picture of the CMB.
Moreover, Large Scale Structure (LSS) surveys by missions such as Euclid~\cite{Amendola:2016saw} can also break $A_s$ and $\taureio$ degeneracy by providing a more precise constraint on $A_s$.

\item The precise astrophysical determination of $\taureio$ is still under development, relying on multiple surveys across a range of $z$. The validity of the conclusion also depends on extrapolations of known results. With further observations from JWST, the preferred range of $\taureio$ may change. Moreover, parallel approaches such as Lyman-$\alpha$ forest~\cite{Ouchi:2020zce,2024MNRAS.528.7052C,2024ApJ...967...28N,Lu:2023rwr} and 21-cm tomography~\cite{2010ARA&A..48..127M,HERA:2021noe} could reveal the history of reionization, further examining the consistency of cosmology.
\end{itemize}

Our results have explored the relationship between two important parameters of cosmology, both studied actively by many collaborations. Future observations of the late universe may lead to a clearer prediction of $\taureio$. In concert, upcoming CMB polarization experiments will provide the most precise inferences of $\taureio$ possible. As such, we have provided a road map for understanding the impact of $\taureio$ on $H_0$ and for interpreting any further shifts in measurements of reionization or the Hubble constant from observations outside of the CMB.

\section*{Acknowledgements}

We thank Nils Sch\"{o}neberg, Hongwan Liu, and Julian Mu\~{n}oz for useful discussions. IJA, JF, LL and PS are supported by the NASA grant 80NSSC22K081 and the DOE grant DE-SC-0010010.  This work was conducted using computational resources and services at the Center for Computation and Visualization, Brown University.  
 
\bibliographystyle{JHEP}
\bibliography{Ref}

\providecommand{\href}[2]{#2}\begingroup\raggedright\begin{thebibliography}{10}

\bibitem{2022A&A...661A..80J}
P.~{Jakobsen}, P.~{Ferruit}, C.~{Alves de Oliveira}, S.~{Arribas},
  G.~{Bagnasco}, R.~{Barho} et~al., \emph{{The Near-Infrared Spectrograph
  (NIRSpec) on the James Webb Space Telescope. I. Overview of the instrument
  and its capabilities}},
  \href{https://doi.org/10.1051/0004-6361/202142663}{\emph{Astron. Astrophys.}
  {\bfseries 661} (2022) A80}
  [\href{https://arxiv.org/abs/2202.03305}{{\ttfamily 2202.03305}}].

\bibitem{2023PASP..135d8003W}
G.S.~{Wright}, G.H.~{Rieke}, A.~{Glasse}, M.~{Ressler}, M.~{Garc{\'\i}a
  Mar{\'\i}n}, J.~{Aguilar} et~al., \emph{{The Mid-infrared Instrument for JWST
  and Its In-flight Performance}},
  \href{https://doi.org/10.1088/1538-3873/acbe66}{\emph{Publ. Astron. Soc.
  Pac.} {\bfseries 135} (2023) 048003}.

\bibitem{Planck:2018vyg}
{\scshape Planck} collaboration, \emph{{Planck 2018 results. VI. Cosmological
  parameters}},
  \href{https://doi.org/10.1051/0004-6361/201833910}{\emph{Astron. Astrophys.}
  {\bfseries 641} (2020) A6}
  [\href{https://arxiv.org/abs/1807.06209}{{\ttfamily 1807.06209}}].

\bibitem{Robertson:2015uda}
B.E.~Robertson, R.S.~Ellis, S.R.~Furlanetto and J.S.~Dunlop, \emph{{Cosmic
  Reionization and Early Star-forming Galaxies: a Joint Analysis of new
  Constraints From Planck and the Hubble Space Telescope}},
  \href{https://doi.org/10.1088/2041-8205/802/2/L19}{\emph{Astrophys. J. Lett.}
  {\bfseries 802} (2015) L19}
  [\href{https://arxiv.org/abs/1502.02024}{{\ttfamily 1502.02024}}].

\bibitem{McQuinn:2015icp}
M.~McQuinn, \emph{{The Evolution of the Intergalactic Medium}},
  \href{https://doi.org/10.1146/annurev-astro-082214-122355}{\emph{Ann. Rev.
  Astron. Astrophys.} {\bfseries 54} (2016) 313}
  [\href{https://arxiv.org/abs/1512.00086}{{\ttfamily 1512.00086}}].

\bibitem{Munoz:2024fas}
J.B.~Mu\~noz, J.~Mirocha, J.~Chisholm, S.R.~Furlanetto and C.~Mason,
  \emph{{Reionization after JWST: a photon budget crisis?}},
  \href{https://doi.org/10.1093/mnrasl/slae086}{\emph{Mon. Not. Roy. Astron.
  Soc.} {\bfseries 535} (2024) L37}
  [\href{https://arxiv.org/abs/2404.07250}{{\ttfamily 2404.07250}}].

\bibitem{2024MNRAS.527.6139S}
C.~{Simmonds}, S.~{Tacchella}, K.~{Hainline}, B.D.~{Johnson}, W.~{McClymont},
  B.~{Robertson} et~al., \emph{{Low-mass bursty galaxies in JADES efficiently
  produce ionizing photons and could represent the main drivers of
  reionization}}, \href{https://doi.org/10.1093/mnras/stad3605}{\emph{Mon. Not.
  R. Astron. Soc.} {\bfseries 527} (2024) 6139}
  [\href{https://arxiv.org/abs/2310.01112}{{\ttfamily 2310.01112}}].

\bibitem{2024MNRAS.533.1111E}
R.~{Endsley}, D.P.~{Stark}, L.~{Whitler}, M.W.~{Topping}, B.D.~{Johnson},
  B.~{Robertson} et~al., \emph{{The star-forming and ionizing properties of
  dwarf z 6-9 galaxies in JADES: insights on bursty star formation and ionized
  bubble growth}}, \href{https://doi.org/10.1093/mnras/stae1857}{\emph{Mon.
  Not. R. Astron. Soc.} {\bfseries 533} (2024) 1111}
  [\href{https://arxiv.org/abs/2306.05295}{{\ttfamily 2306.05295}}].

\bibitem{2023ApJ...946L..13F}
S.L.~{Finkelstein}, M.B.~{Bagley}, H.C.~{Ferguson}, S.M.~{Wilkins},
  J.S.~{Kartaltepe}, C.~{Papovich} et~al., \emph{{CEERS Key Paper. I. An Early
  Look into the First 500 Myr of Galaxy Formation with JWST}},
  \href{https://doi.org/10.3847/2041-8213/acade4}{\emph{Astrophys. J. Lett.}
  {\bfseries 946} (2023) L13}
  [\href{https://arxiv.org/abs/2211.05792}{{\ttfamily 2211.05792}}].

\bibitem{2024ApJ...969L...2F}
S.L.~{Finkelstein}, G.C.K.~{Leung}, M.B.~{Bagley}, M.~{Dickinson},
  H.C.~{Ferguson}, C.~{Papovich} et~al., \emph{{The Complete CEERS Early
  Universe Galaxy Sample: A Surprisingly Slow Evolution of the Space Density of
  Bright Galaxies at z {\ensuremath{\sim}} 8.5{\textendash}14.5}},
  \href{https://doi.org/10.3847/2041-8213/ad4495}{\emph{Astrophys. J. Lett.}
  {\bfseries 969} (2024) L2}
  [\href{https://arxiv.org/abs/2311.04279}{{\ttfamily 2311.04279}}].

\bibitem{2022MNRAS.517.5104C}
J.~{Chisholm}, A.~{Saldana-Lopez}, S.~{Flury}, D.~{Schaerer}, A.~{Jaskot},
  R.~{Amor{\'\i}n} et~al., \emph{{The far-ultraviolet continuum slope as a
  Lyman Continuum escape estimator at high redshift}},
  \href{https://doi.org/10.1093/mnras/stac2874}{\emph{Mon. Not. R. Astron.
  Soc.} {\bfseries 517} (2022) 5104}
  [\href{https://arxiv.org/abs/2207.05771}{{\ttfamily 2207.05771}}].

\bibitem{Mukherjee:2024cfq}
P.~Mukherjee, A.~Dey and S.~Pal, \emph{{What can we learn about Reionization
  astrophysical parameters using Gaussian Process Regression?}},
  \href{https://arxiv.org/abs/2407.19481}{{\ttfamily 2407.19481}}.

\bibitem{Paoletti:2024lji}
D.~Paoletti, D.K.~Hazra, F.~Finelli and G.F.~Smoot, \emph{{The asymmetry of
  dawn: evidence for asymmetric reionization histories from a joint analysis of
  cosmic microwave background and astrophysical data}},
  \href{https://arxiv.org/abs/2405.09506}{{\ttfamily 2405.09506}}.

\bibitem{Zhu:2024xrt}
Y.~Zhu et~al., \emph{{SMILES: Discovery of Higher Ionizing Photon Production
  Efficiency in Overdense Regions}},
  \href{https://arxiv.org/abs/2410.14804}{{\ttfamily 2410.14804}}.

\bibitem{Cain:2024fbi}
C.~Cain, G.~Lopez, A.~D'Aloisio, J.B.~Munoz, R.A.~Jansen, R.A.~Windhorst
  et~al., \emph{{Chasing the beginning of reionization in the JWST era}},
  \href{https://arxiv.org/abs/2409.02989}{{\ttfamily 2409.02989}}.

\bibitem{Giare:2023ejv}
W.~Giar\`e, E.~Di~Valentino and A.~Melchiorri, \emph{{Measuring the
  reionization optical depth without large-scale CMB polarization}},
  \href{https://doi.org/10.1103/PhysRevD.109.103519}{\emph{Phys. Rev. D}
  {\bfseries 109} (2024) 103519}
  [\href{https://arxiv.org/abs/2312.06482}{{\ttfamily 2312.06482}}].

\bibitem{Riess:2021jrx}
A.G.~Riess et~al., \emph{{A Comprehensive Measurement of the Local Value of the
  Hubble Constant with 1 km s$^{-1}$ Mpc$^{-1}$ Uncertainty from the Hubble
  Space Telescope and the SH0ES Team}},
  \href{https://doi.org/10.3847/2041-8213/ac5c5b}{\emph{Astrophys. J. Lett.}
  {\bfseries 934} (2022) L7}
  [\href{https://arxiv.org/abs/2112.04510}{{\ttfamily 2112.04510}}].

\bibitem{Breuval:2024lsv}
L.~Breuval, A.G.~Riess, S.~Casertano, W.~Yuan, L.M.~Macri, M.~Romaniello
  et~al., \emph{{Small Magellanic Cloud Cepheids Observed with the Hubble Space
  Telescope Provide a New Anchor for the SH0ES Distance Ladder}},
  \href{https://doi.org/10.3847/1538-4357/ad630e}{\emph{Astrophys. J.}
  {\bfseries 973} (2024) 30}
  [\href{https://arxiv.org/abs/2404.08038}{{\ttfamily 2404.08038}}].

\bibitem{Scolnic:2023mrv}
D.~Scolnic, A.G.~Riess, J.~Wu, S.~Li, G.S.~Anand, R.~Beaton et~al.,
  \emph{{CATS: The Hubble Constant from Standardized TRGB and Type Ia Supernova
  Measurements}},
  \href{https://doi.org/10.3847/2041-8213/ace978}{\emph{Astrophys. J. Lett.}
  {\bfseries 954} (2023) L31}
  [\href{https://arxiv.org/abs/2304.06693}{{\ttfamily 2304.06693}}].

\bibitem{Riess:2024vfa}
A.G.~Riess et~al., \emph{{JWST Validates HST Distance Measurements: Selection
  of Supernova Subsample Explains Differences in JWST Estimates of Local H
  $_{0}$}}, \href{https://doi.org/10.3847/1538-4357/ad8c21}{\emph{Astrophys.
  J.} {\bfseries 977} (2024) 120}
  [\href{https://arxiv.org/abs/2408.11770}{{\ttfamily 2408.11770}}].

\bibitem{H0LiCOW:2019pvv}
{\scshape H0LiCOW} collaboration, \emph{{H0LiCOW \textendash{} XIII. A 2.4 per
  cent measurement of H0 from lensed quasars: 5.3\ensuremath{\sigma} tension
  between early- and late-Universe probes}},
  \href{https://doi.org/10.1093/mnras/stz3094}{\emph{Mon. Not. Roy. Astron.
  Soc.} {\bfseries 498} (2020) 1420}
  [\href{https://arxiv.org/abs/1907.04869}{{\ttfamily 1907.04869}}].

\bibitem{Freedman:2024eph}
W.L.~Freedman, B.F.~Madore, I.S.~Jang, T.J.~Hoyt, A.J.~Lee and K.A.~Owens,
  \emph{{Status Report on the Chicago-Carnegie Hubble Program (CCHP): Three
  Independent Astrophysical Determinations of the Hubble Constant Using the
  James Webb Space Telescope}},
  \href{https://arxiv.org/abs/2408.06153}{{\ttfamily 2408.06153}}.

\bibitem{Freedman:2023jcz}
W.L.~Freedman and B.F.~Madore, \emph{{Progress in direct measurements of the
  Hubble constant}},
  \href{https://doi.org/10.1088/1475-7516/2023/11/050}{\emph{JCAP} {\bfseries
  11} (2023) 050} [\href{https://arxiv.org/abs/2309.05618}{{\ttfamily
  2309.05618}}].

\bibitem{Freedman:2021ahq}
W.L.~Freedman, \emph{{Measurements of the Hubble Constant: Tensions in
  Perspective}},
  \href{https://doi.org/10.3847/1538-4357/ac0e95}{\emph{Astrophys. J.}
  {\bfseries 919} (2021) 16}
  [\href{https://arxiv.org/abs/2106.15656}{{\ttfamily 2106.15656}}].

\bibitem{Archidiacono:2013fha}
M.~Archidiacono, E.~Giusarma, S.~Hannestad and O.~Mena, \emph{{Cosmic dark
  radiation and neutrinos}},
  \href{https://doi.org/10.1155/2013/191047}{\emph{Adv. High Energy Phys.}
  {\bfseries 2013} (2013) 191047}
  [\href{https://arxiv.org/abs/1307.0637}{{\ttfamily 1307.0637}}].

\bibitem{Akita:2020szl}
K.~Akita and M.~Yamaguchi, \emph{{A precision calculation of relic neutrino
  decoupling}},
  \href{https://doi.org/10.1088/1475-7516/2020/08/012}{\emph{JCAP} {\bfseries
  08} (2020) 012} [\href{https://arxiv.org/abs/2005.07047}{{\ttfamily
  2005.07047}}].

\bibitem{Froustey:2020mcq}
J.~Froustey, C.~Pitrou and M.C.~Volpe, \emph{{Neutrino decoupling including
  flavour oscillations and primordial nucleosynthesis}},
  \href{https://doi.org/10.1088/1475-7516/2020/12/015}{\emph{JCAP} {\bfseries
  12} (2020) 015} [\href{https://arxiv.org/abs/2008.01074}{{\ttfamily
  2008.01074}}].

\bibitem{Bennett:2020zkv}
J.J.~Bennett, G.~Buldgen, P.F.~De~Salas, M.~Drewes, S.~Gariazzo, S.~Pastor
  et~al., \emph{{Towards a precision calculation of $N_{\rm eff}$ in the
  Standard Model II: Neutrino decoupling in the presence of flavour
  oscillations and finite-temperature QED}},
  \href{https://doi.org/10.1088/1475-7516/2021/04/073}{\emph{JCAP} {\bfseries
  04} (2021) 073} [\href{https://arxiv.org/abs/2012.02726}{{\ttfamily
  2012.02726}}].

\bibitem{DiValentino:2021izs}
E.~Di~Valentino, O.~Mena, S.~Pan, L.~Visinelli, W.~Yang, A.~Melchiorri et~al.,
  \emph{{In the realm of the Hubble tension\textemdash{}a review of
  solutions}}, \href{https://doi.org/10.1088/1361-6382/ac086d}{\emph{Class.
  Quant. Grav.} {\bfseries 38} (2021) 153001}
  [\href{https://arxiv.org/abs/2103.01183}{{\ttfamily 2103.01183}}].

\bibitem{Schoneberg:2021qvd}
N.~Sch\"oneberg, G.~Franco~Abell\'an, A.~P\'erez~S\'anchez, S.J.~Witte,
  V.~Poulin and J.~Lesgourgues, \emph{{The H0 Olympics: A fair ranking of
  proposed models}},
  \href{https://doi.org/10.1016/j.physrep.2022.07.001}{\emph{Phys. Rept.}
  {\bfseries 984} (2022) 1} [\href{https://arxiv.org/abs/2107.10291}{{\ttfamily
  2107.10291}}].

\bibitem{Jeong:2013eza}
K.S.~Jeong and F.~Takahashi, \emph{{Self-interacting Dark Radiation}},
  \href{https://doi.org/10.1016/j.physletb.2013.07.001}{\emph{Phys. Lett. B}
  {\bfseries 725} (2013) 134}
  [\href{https://arxiv.org/abs/1305.6521}{{\ttfamily 1305.6521}}].

\bibitem{Buen-Abad:2015ova}
M.A.~Buen-Abad, G.~Marques-Tavares and M.~Schmaltz, \emph{{Non-Abelian dark
  matter and dark radiation}},
  \href{https://doi.org/10.1103/PhysRevD.92.023531}{\emph{Phys. Rev. D}
  {\bfseries 92} (2015) 023531}
  [\href{https://arxiv.org/abs/1505.03542}{{\ttfamily 1505.03542}}].

\bibitem{Buen-Abad:2017gxg}
M.A.~Buen-Abad, M.~Schmaltz, J.~Lesgourgues and T.~Brinckmann,
  \emph{{Interacting Dark Sector and Precision Cosmology}},
  \href{https://doi.org/10.1088/1475-7516/2018/01/008}{\emph{JCAP} {\bfseries
  01} (2018) 008} [\href{https://arxiv.org/abs/1708.09406}{{\ttfamily
  1708.09406}}].

\bibitem{Brust:2017nmv}
C.~Brust, Y.~Cui and K.~Sigurdson, \emph{{Cosmological Constraints on
  Interacting Light Particles}},
  \href{https://doi.org/10.1088/1475-7516/2017/08/020}{\emph{JCAP} {\bfseries
  08} (2017) 020} [\href{https://arxiv.org/abs/1703.10732}{{\ttfamily
  1703.10732}}].

\bibitem{Blinov:2020hmc}
N.~Blinov and G.~Marques-Tavares, \emph{{Interacting radiation after Planck and
  its implications for the Hubble Tension}},
  \href{https://doi.org/10.1088/1475-7516/2020/09/029}{\emph{JCAP} {\bfseries
  09} (2020) 029} [\href{https://arxiv.org/abs/2003.08387}{{\ttfamily
  2003.08387}}].

\bibitem{Brinckmann:2022ajr}
T.~Brinckmann, J.H.~Chang, P.~Du and M.~LoVerde, \emph{{Confronting interacting
  dark radiation scenarios with cosmological data}},
  \href{https://doi.org/10.1103/PhysRevD.107.123517}{\emph{Phys. Rev. D}
  {\bfseries 107} (2023) 123517}
  [\href{https://arxiv.org/abs/2212.13264}{{\ttfamily 2212.13264}}].

\bibitem{Buen-Abad:2024tlb}
M.A.~Buen-Abad, Z.~Chacko, I.~Flood, C.~Kilic, G.~Marques-Tavares and T.~Youn,
  \emph{{Atomic Dark Matter, Interacting Dark Radiation, and the Hubble
  Tension}},  \href{https://arxiv.org/abs/2411.08097}{{\ttfamily 2411.08097}}.

\bibitem{Karwal:2016vyq}
T.~Karwal and M.~Kamionkowski, \emph{{Dark energy at early times, the Hubble
  parameter, and the string axiverse}},
  \href{https://doi.org/10.1103/PhysRevD.94.103523}{\emph{Phys. Rev. D}
  {\bfseries 94} (2016) 103523}
  [\href{https://arxiv.org/abs/1608.01309}{{\ttfamily 1608.01309}}].

\bibitem{Poulin:2018cxd}
V.~Poulin, T.L.~Smith, T.~Karwal and M.~Kamionkowski, \emph{{Early Dark Energy
  Can Resolve The Hubble Tension}},
  \href{https://doi.org/10.1103/PhysRevLett.122.221301}{\emph{Phys. Rev. Lett.}
  {\bfseries 122} (2019) 221301}
  [\href{https://arxiv.org/abs/1811.04083}{{\ttfamily 1811.04083}}].

\bibitem{Smith_2020}
T.L.~Smith, V.~Poulin and M.A.~Amin, \emph{Oscillating scalar fields and the
  hubble tension: A resolution with novel signatures},
  \href{https://doi.org/10.1103/physrevd.101.063523}{\emph{Physical Review D}
  {\bfseries 101} (2020) }.

\bibitem{Diego_Blas_2011}
D.~Blas, J.~Lesgourgues and T.~Tram, \emph{The cosmic linear anisotropy solving
  system (class). part ii: Approximation schemes},
  \href{https://doi.org/10.1088/1475-7516/2011/07/034}{\emph{JCAP} {\bfseries
  2011} (2011) 034–034}.

\bibitem{lesgourgues2011cosmiclinearanisotropysolving}
J.~Lesgourgues, \emph{The cosmic linear anisotropy solving system (class) i:
  Overview},  2011.

\bibitem{Smith:2019ihp}
T.L.~Smith, V.~Poulin and M.A.~Amin, \emph{{Oscillating scalar fields and the
  Hubble tension: a resolution with novel signatures}},
  \href{https://doi.org/10.1103/PhysRevD.101.063523}{\emph{Phys. Rev. D}
  {\bfseries 101} (2020) 063523}
  [\href{https://arxiv.org/abs/1908.06995}{{\ttfamily 1908.06995}}].

\bibitem{Poulin:2018dzj}
V.~Poulin, T.L.~Smith, D.~Grin, T.~Karwal and M.~Kamionkowski,
  \emph{{Cosmological implications of ultralight axionlike fields}},
  \href{https://doi.org/10.1103/PhysRevD.98.083525}{\emph{Phys. Rev. D}
  {\bfseries 98} (2018) 083525}
  [\href{https://arxiv.org/abs/1806.10608}{{\ttfamily 1806.10608}}].

\bibitem{2019ascl.soft10019T}
J.~{Torrado} and A.~{Lewis}, ``{Cobaya: Bayesian analysis in cosmology}.''
  Astrophysics Source Code Library, record ascl:1910.019, Oct., 2019.

\bibitem{Torrado:2020dgo}
J.~Torrado and A.~Lewis, \emph{{Cobaya: Code for Bayesian Analysis of
  hierarchical physical models}},
  \href{https://doi.org/10.1088/1475-7516/2021/05/057}{\emph{JCAP} {\bfseries
  05} (2021) 057} [\href{https://arxiv.org/abs/2005.05290}{{\ttfamily
  2005.05290}}].

\bibitem{lewis2019getdistpythonpackageanalysing}
A.~Lewis, \emph{Getdist: a python package for analysing monte carlo samples},
  2019.

\bibitem{Planck:2019nip}
{\scshape Planck} collaboration, \emph{{Planck 2018 results. V. CMB power
  spectra and likelihoods}},
  \href{https://doi.org/10.1051/0004-6361/201936386}{\emph{Astron. Astrophys.}
  {\bfseries 641} (2020) A5}
  [\href{https://arxiv.org/abs/1907.12875}{{\ttfamily 1907.12875}}].

\bibitem{Efstathiou_2021}
G.~Efstathiou and S.~Gratton, \emph{A detailed description of the camspec
  likelihood pipeline and a reanalysis of the planck high frequency maps},
  \href{https://doi.org/10.21105/astro.1910.00483}{\emph{The Open Journal of
  Astrophysics} {\bfseries 4} (2021) }.

\bibitem{Planck:2020olo}
{\scshape Planck} collaboration, \emph{{$Planck$ intermediate results. LVII.
  Joint Planck LFI and HFI data processing}},
  \href{https://doi.org/10.1051/0004-6361/202038073}{\emph{Astron. Astrophys.}
  {\bfseries 643} (2020) A42}
  [\href{https://arxiv.org/abs/2007.04997}{{\ttfamily 2007.04997}}].

\bibitem{Rosenberg_2022}
E.~Rosenberg, S.~Gratton and G.~Efstathiou, \emph{Cmb power spectra and
  cosmological parameters from planck pr4 with camspec},
  \href{https://doi.org/10.1093/mnras/stac2744}{\emph{Mon. Not. Roy. Astron.
  Soc.} {\bfseries 517} (2022) 4620–4636}.

\bibitem{ACT:2023dou}
{\scshape ACT} collaboration, \emph{{The Atacama Cosmology Telescope: A
  Measurement of the DR6 CMB Lensing Power Spectrum and Its Implications for
  Structure Growth}},
  \href{https://doi.org/10.3847/1538-4357/acfe06}{\emph{Astrophys. J.}
  {\bfseries 962} (2024) 112}
  [\href{https://arxiv.org/abs/2304.05202}{{\ttfamily 2304.05202}}].

\bibitem{ACT:2023kun}
{\scshape ACT} collaboration, \emph{{The Atacama Cosmology Telescope: DR6
  Gravitational Lensing Map and Cosmological Parameters}},
  \href{https://doi.org/10.3847/1538-4357/acff5f}{\emph{Astrophys. J.}
  {\bfseries 962} (2024) 113}
  [\href{https://arxiv.org/abs/2304.05203}{{\ttfamily 2304.05203}}].

\bibitem{Carron:2022eyg}
J.~Carron, M.~Mirmelstein and A.~Lewis, \emph{{CMB lensing from Planck
  PR4~maps}}, \href{https://doi.org/10.1088/1475-7516/2022/09/039}{\emph{JCAP}
  {\bfseries 09} (2022) 039}
  [\href{https://arxiv.org/abs/2206.07773}{{\ttfamily 2206.07773}}].

\bibitem{Scolnic:2021amr}
D.~Scolnic et~al., \emph{{The Pantheon+ Analysis: The Full Data Set and
  Light-curve Release}},
  \href{https://doi.org/10.3847/1538-4357/ac8b7a}{\emph{Astrophys. J.}
  {\bfseries 938} (2022) 113}
  [\href{https://arxiv.org/abs/2112.03863}{{\ttfamily 2112.03863}}].

\bibitem{Beutler_2011}
F.~Beutler, C.~Blake, M.~Colless, D.H.~Jones, L.~Staveley-Smith, L.~Campbell
  et~al., \emph{The 6df galaxy survey: baryon acoustic oscillations and the
  local hubble constant: 6dfgs: Baos and the local hubble constant},
  \href{https://doi.org/10.1111/j.1365-2966.2011.19250.x}{\emph{Mon. Not. Roy.
  Astron. Soc.} {\bfseries 416} (2011) 3017–3032}.

\bibitem{Ross:2014qpa}
A.J.~Ross, L.~Samushia, C.~Howlett, W.J.~Percival, A.~Burden and M.~Manera,
  \emph{{The clustering of the SDSS DR7 main Galaxy sample \textendash{} I. A 4
  per cent distance measure at $z = 0.15$}},
  \href{https://doi.org/10.1093/mnras/stv154}{\emph{Mon. Not. Roy. Astron.
  Soc.} {\bfseries 449} (2015) 835}
  [\href{https://arxiv.org/abs/1409.3242}{{\ttfamily 1409.3242}}].

\bibitem{BOSS:2016wmc}
{\scshape BOSS} collaboration, \emph{{The clustering of galaxies in the
  completed SDSS-III Baryon Oscillation Spectroscopic Survey: cosmological
  analysis of the DR12 galaxy sample}},
  \href{https://doi.org/10.1093/mnras/stx721}{\emph{Mon. Not. Roy. Astron.
  Soc.} {\bfseries 470} (2017) 2617}
  [\href{https://arxiv.org/abs/1607.03155}{{\ttfamily 1607.03155}}].

\bibitem{DESI:2024mwx}
{\scshape DESI} collaboration, \emph{{DESI 2024 VI: Cosmological Constraints
  from the Measurements of Baryon Acoustic Oscillations}},
  \href{https://arxiv.org/abs/2404.03002}{{\ttfamily 2404.03002}}.

\bibitem{Allali:2024cji}
I.J.~Allali, A.~Notari and F.~Rompineve, \emph{{Dark Radiation with Baryon
  Acoustic Oscillations from DESI 2024 and the $H_0$ tension}},
  \href{https://arxiv.org/abs/2404.15220}{{\ttfamily 2404.15220}}.

\bibitem{Qu:2024lpx}
F.J.~Qu, K.M.~Surrao, B.~Bolliet, J.C.~Hill, B.D.~Sherwin and H.T.~Jense,
  \emph{{Accelerated inference on accelerated cosmic expansion: New constraints
  on axion-like early dark energy with DESI BAO and ACT DR6 CMB lensing}},
  \href{https://arxiv.org/abs/2404.16805}{{\ttfamily 2404.16805}}.

\bibitem{Wang:2024dka}
H.~Wang and Y.-S.~Piao, \emph{{Dark energy in light of recent DESI BAO and
  Hubble tension}},  \href{https://arxiv.org/abs/2404.18579}{{\ttfamily
  2404.18579}}.

\bibitem{Seto:2024cgo}
O.~Seto and Y.~Toda, \emph{{DESI constraints on the varying electron mass model
  and axionlike early dark energy}},
  \href{https://doi.org/10.1103/PhysRevD.110.083501}{\emph{Phys. Rev. D}
  {\bfseries 110} (2024) 083501}
  [\href{https://arxiv.org/abs/2405.11869}{{\ttfamily 2405.11869}}].

\bibitem{Lynch:2024hzh}
G.P.~Lynch, L.~Knox and J.~Chluba, \emph{{DESI observations and the Hubble
  tension in light of modified recombination}},
  \href{https://doi.org/10.1103/PhysRevD.110.083538}{\emph{Phys. Rev. D}
  {\bfseries 110} (2024) 083538}
  [\href{https://arxiv.org/abs/2406.10202}{{\ttfamily 2406.10202}}].

\bibitem{Jiang:2024xnu}
J.-Q.~Jiang, D.~Pedrotti, S.S.~da~Costa and S.~Vagnozzi, \emph{{Non-parametric
  late-time expansion history reconstruction and implications for the Hubble
  tension in light of recent DESI and Type Ia supernovae data}},
  \href{https://arxiv.org/abs/2408.02365}{{\ttfamily 2408.02365}}.

\bibitem{Chatrchyan:2024xjj}
A.~Chatrchyan, F.~Niedermann, V.~Poulin and M.S.~Sloth, \emph{{Confronting cold
  new early dark energy and its equation of state with updated CMB, supernovae,
  and BAO data}},
  \href{https://doi.org/10.1103/PhysRevD.111.043536}{\emph{Phys. Rev. D}
  {\bfseries 111} (2025) 043536}
  [\href{https://arxiv.org/abs/2408.14537}{{\ttfamily 2408.14537}}].

\bibitem{Li:2025rjr}
Y.-Z.~Li and J.-H.~Yu, \emph{{Neutrinophilic $\mathbf{\Lambda}$CDM Extension
  for EMPRESS, DESI and Hubble Tension}},
  \href{https://arxiv.org/abs/2501.13153}{{\ttfamily 2501.13153}}.

\bibitem{Lesgourgues:2013qba}
J.~Lesgourgues, \emph{{Cosmological Perturbations}},  in \emph{{Theoretical
  Advanced Study Institute in Elementary Particle Physics}: {Searching for New
  Physics at Small and Large Scales}}, pp.~29--97, 2013,
  \href{https://doi.org/10.1142/9789814525220_0002}{DOI}
  [\href{https://arxiv.org/abs/1302.4640}{{\ttfamily 1302.4640}}].

\bibitem{LiteBIRD:2020khw}
{\scshape LiteBIRD} collaboration, \emph{{LiteBIRD: JAXA's new strategic
  L-class mission for all-sky surveys of cosmic microwave background
  polarization}}, \href{https://doi.org/10.1117/12.2563050}{\emph{Proc. SPIE
  Int. Soc. Opt. Eng.} {\bfseries 11443} (2020) 114432F}
  [\href{https://arxiv.org/abs/2101.12449}{{\ttfamily 2101.12449}}].

\bibitem{Sen:2022usj}
A.~Sen, S.~Basak, T.~Ghosh, D.~Adak and S.~Sinha, \emph{{Importance of
  high-frequency bands for removal of thermal dust in ECHO}},
  \href{https://doi.org/10.1103/PhysRevD.108.083529}{\emph{Phys. Rev. D}
  {\bfseries 108} (2023) 083529}
  [\href{https://arxiv.org/abs/2212.02869}{{\ttfamily 2212.02869}}].

\bibitem{Adak:2021lbu}
D.~Adak, A.~Sen, S.~Basak, J.~Delabrouille, T.~Ghosh, A.~Rotti et~al.,
  \emph{{B-mode forecast of CMB-Bh\={a}rat}},
  \href{https://doi.org/10.1093/mnras/stac1474}{\emph{Mon. Not. Roy. Astron.
  Soc.} {\bfseries 514} (2022) 3002}
  [\href{https://arxiv.org/abs/2110.12362}{{\ttfamily 2110.12362}}].

\bibitem{PICO:2019thw}
{\scshape NASA PICO} collaboration, \emph{{PICO: Probe of Inflation and Cosmic
  Origins}},  \href{https://arxiv.org/abs/1902.10541}{{\ttfamily 1902.10541}}.

\bibitem{SimonsObservatory:2018koc}
{\scshape Simons Observatory} collaboration, \emph{{The Simons Observatory:
  Science goals and forecasts}},
  \href{https://doi.org/10.1088/1475-7516/2019/02/056}{\emph{JCAP} {\bfseries
  02} (2019) 056} [\href{https://arxiv.org/abs/1808.07445}{{\ttfamily
  1808.07445}}].

\bibitem{POLARBEAR:2016wwl}
{\scshape POLARBEAR} collaboration, \emph{{POLARBEAR-2: an instrument for CMB
  polarization measurements}},
  \href{https://doi.org/10.1117/12.2231961}{\emph{Proc. SPIE Int. Soc. Opt.
  Eng.} {\bfseries 9914} (2016) 99141I}
  [\href{https://arxiv.org/abs/1608.03025}{{\ttfamily 1608.03025}}].

\bibitem{Bicep}
L.~Moncelsi et~al., \emph{{Receiver development for BICEP Array, a
  next-generation CMB polarimeter at the South Pole}},
  \href{https://doi.org/10.1117/12.2561995}{\emph{Proc. SPIE Int. Soc. Opt.
  Eng.} {\bfseries 11453} (2020) 1145314}
  [\href{https://arxiv.org/abs/2012.04047}{{\ttfamily 2012.04047}}].

\bibitem{Amendola:2016saw}
L.~Amendola et~al., \emph{{Cosmology and fundamental physics with the Euclid
  satellite}}, \href{https://doi.org/10.1007/s41114-017-0010-3}{\emph{Living
  Rev. Rel.} {\bfseries 21} (2018) 2}
  [\href{https://arxiv.org/abs/1606.00180}{{\ttfamily 1606.00180}}].

\bibitem{Ouchi:2020zce}
M.~Ouchi, Y.~Ono and T.~Shibuya, \emph{{Observations of the Lyman-$\alpha$
  Universe}},
  \href{https://doi.org/10.1146/annurev-astro-032620-021859}{\emph{Ann. Rev.
  Astron. Astrophys.} {\bfseries 58} (2020) 617}
  [\href{https://arxiv.org/abs/2012.07960}{{\ttfamily 2012.07960}}].

\bibitem{2024MNRAS.528.7052C}
Z.~{Chen}, D.P.~{Stark}, C.~{Mason}, M.W.~{Topping}, L.~{Whitler}, M.~{Tang}
  et~al., \emph{{JWST spectroscopy of z 5-8 UV-selected galaxies: new
  constraints on the evolution of the Ly {\ensuremath{\alpha}} escape fraction
  in the reionization era}},
  \href{https://doi.org/10.1093/mnras/stae455}{\emph{Mon. Not. R. Astron. Soc.}
  {\bfseries 528} (2024) 7052}
  [\href{https://arxiv.org/abs/2311.13683}{{\ttfamily 2311.13683}}].

\bibitem{2024ApJ...967...28N}
M.~{Nakane}, M.~{Ouchi}, K.~{Nakajima}, Y.~{Harikane}, Y.~{Ono}, H.~{Umeda}
  et~al., \emph{{Ly{\ensuremath{\alpha}} Emission at z = 7{\textendash}13:
  Clear Evolution of Ly{\ensuremath{\alpha}} Equivalent Width Indicating a Late
  Cosmic Reionization History}},
  \href{https://doi.org/10.3847/1538-4357/ad38c2}{\emph{Astrophys. J.}
  {\bfseries 967} (2024) 28}
  [\href{https://arxiv.org/abs/2312.06804}{{\ttfamily 2312.06804}}].

\bibitem{Lu:2023rwr}
T.-Y.~Lu, C.A.~Mason, A.~Hutter, A.~Mesinger, Y.~Qin, D.P.~Stark et~al.,
  \emph{{The reionizing bubble size distribution around galaxies}},
  \href{https://doi.org/10.1093/mnras/stae266}{\emph{Mon. Not. Roy. Astron.
  Soc.} {\bfseries 528} (2024) 4872}
  [\href{https://arxiv.org/abs/2304.11192}{{\ttfamily 2304.11192}}].

\bibitem{2010ARA&A..48..127M}
M.F.~{Morales} and J.S.B.~{Wyithe}, \emph{{Reionization and Cosmology with
  21-cm Fluctuations}},
  \href{https://doi.org/10.1146/annurev-astro-081309-130936}{\emph{Annu. Rev.
  Astron. Astrophys.} {\bfseries 48} (2010) 127}
  [\href{https://arxiv.org/abs/0910.3010}{{\ttfamily 0910.3010}}].

\bibitem{HERA:2021noe}
{\scshape HERA} collaboration, \emph{{HERA Phase I Limits on the Cosmic 21 cm
  Signal: Constraints on Astrophysics and Cosmology during the Epoch of
  Reionization}},
  \href{https://doi.org/10.3847/1538-4357/ac2ffc}{\emph{Astrophys. J.}
  {\bfseries 924} (2022) 51}
  [\href{https://arxiv.org/abs/2108.07282}{{\ttfamily 2108.07282}}].

\bibitem{Kilo-DegreeSurvey:2023gfr}
{\scshape Kilo-Degree Survey, DES} collaboration, \emph{{DES Y3 + KiDS-1000:
  Consistent cosmology combining cosmic shear surveys}},
  \href{https://doi.org/10.21105/astro.2305.17173}{\emph{Open J. Astrophys.}
  {\bfseries 6} (2023) 2305.17173}
  [\href{https://arxiv.org/abs/2305.17173}{{\ttfamily 2305.17173}}].

\end{thebibliography}\endgroup

\appendix

\section{Detailed Posterior Statistics}
\label{appendix}
We provide the posterior distributions of all  cosmological parameters for various models and datasets discussed previously. Additionally, we present analyses including the local measurement of $H_0$, using Cepheid-calibrated supernova distances, by the SH0ES collaboration. For this dataset, which we will denote \shoes, we use the implementation in \texttt{Cobaya} denoted ``pantheonplusshoes" which replaces the Pantheon+ dataset in \baseline~with a new Pantheon+ dataset calibrated using the intrinsic Type Ia supernova brightness $M_b=-19.253 \pm 0.027$, as reported by the SH0ES collaboration~\cite{Riess:2021jrx}.

\begin{figure}[H]
    \centering
    \includegraphics[scale=1.0]{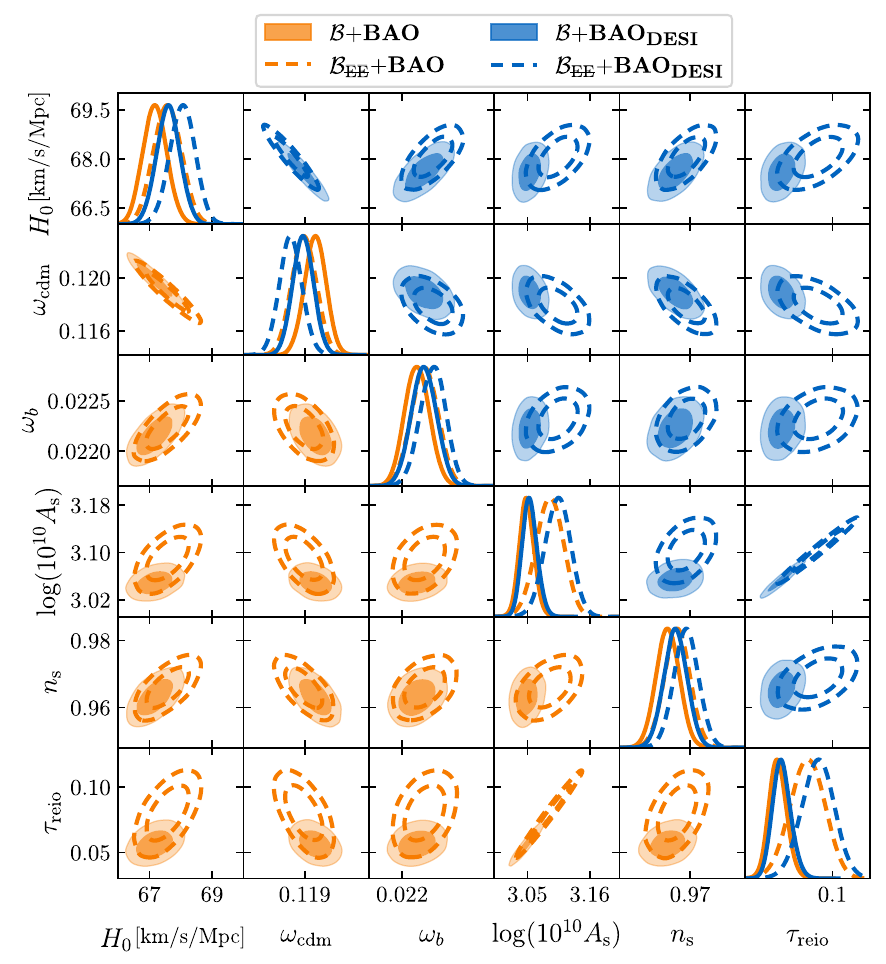}
    \caption{
    Posterior distributions of the cosmological parameters in the $\Lambda$CDM model. The results from the \bao~(\desi) dataset are shown in the lower-left (upper-right) part of the plot as orange (blue) contours. Solid (dashed) contours represent the results obtained from the \baseline~(\baselinewoee) dataset.}
    \label{fig:LCDM_bao_desi}
\end{figure}

\begin{figure}[]
    \centering
    \includegraphics[scale=1.0]{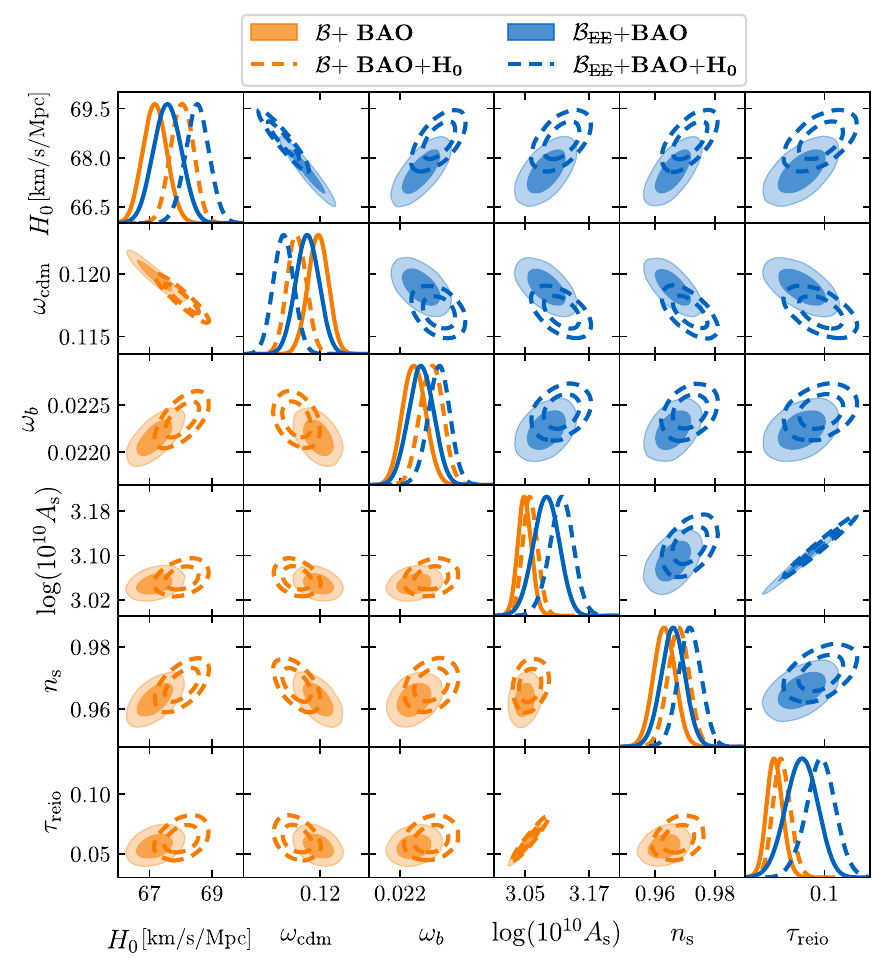}
    \caption{
   Posterior distributions of the cosmological parameters in the $\Lambda$CDM model. The orange (blue) contours show the \baseline+\bao~(\baselinewoee+\bao) dataset. In dashed contours, the \shoes~dataset is included, pulling the posteriors to the high $H_0$ region.}
    \label{fig:LCDM_bao_shoes}
\end{figure}

\begin{figure}[H]
    \centering
    \includegraphics[scale=1.0]{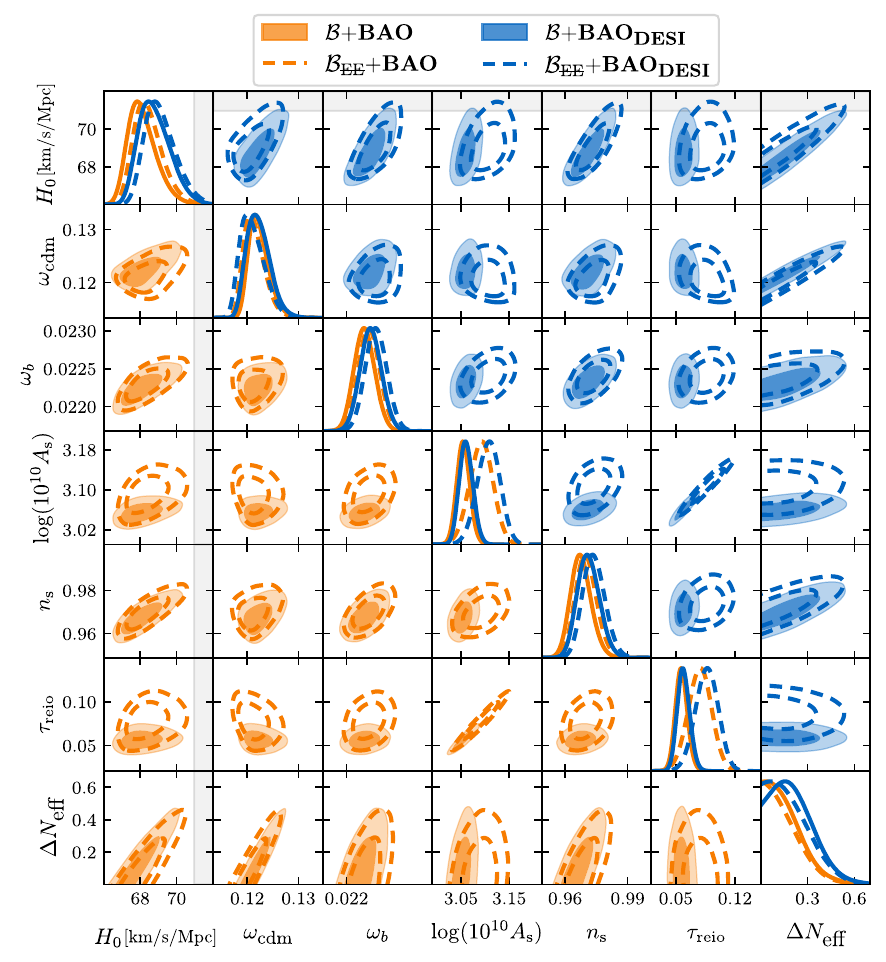}
    \caption{Posterior distributions of the cosmological parameters in the FSDR model along with its new physics parameter, $\DNeff$. The results from the \bao~(\desi) dataset are shown in the lower-left (upper-right) part of the plot as orange (blue) contours.  Solid (dashed) contours represent the results obtained from the \baseline~(\baselinewoee) dataset. The gray shaded area shows the 2$\sigma$ range of $H_0$ from the SH0ES measurement.}
    \label{fig:fsdr_bao_desi}
\end{figure}

\begin{figure}[]
    \centering
    \includegraphics[scale=1.0]{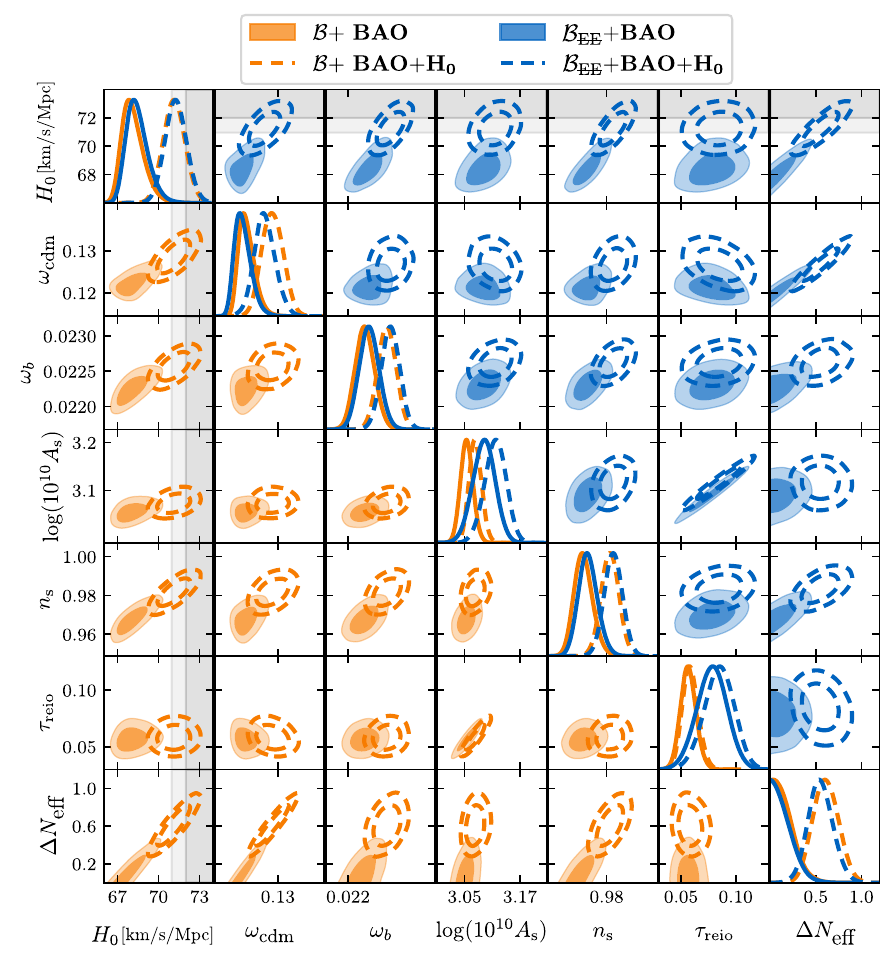}
    \caption{
    Posterior distributions of the cosmological parameters in the FSDR model along with its new physics parameter, $\DNeff$. The orange (blue) contours show the \baseline+\bao~(\baselinewoee+\bao) dataset. 
    In dashed contours, the \shoes~dataset is included, pulling the posteriors to the high $H_0$ region. The gray shaded area shows the $1\sigma$ (dark gray) and $2\sigma$ (light gray) range of $H_0$ from the SH0ES measurement. 
    } 
    \label{fig:fsdr_bao_shoes}
\end{figure}

\begin{figure}[H]
    \centering
    \includegraphics[scale=1.0]{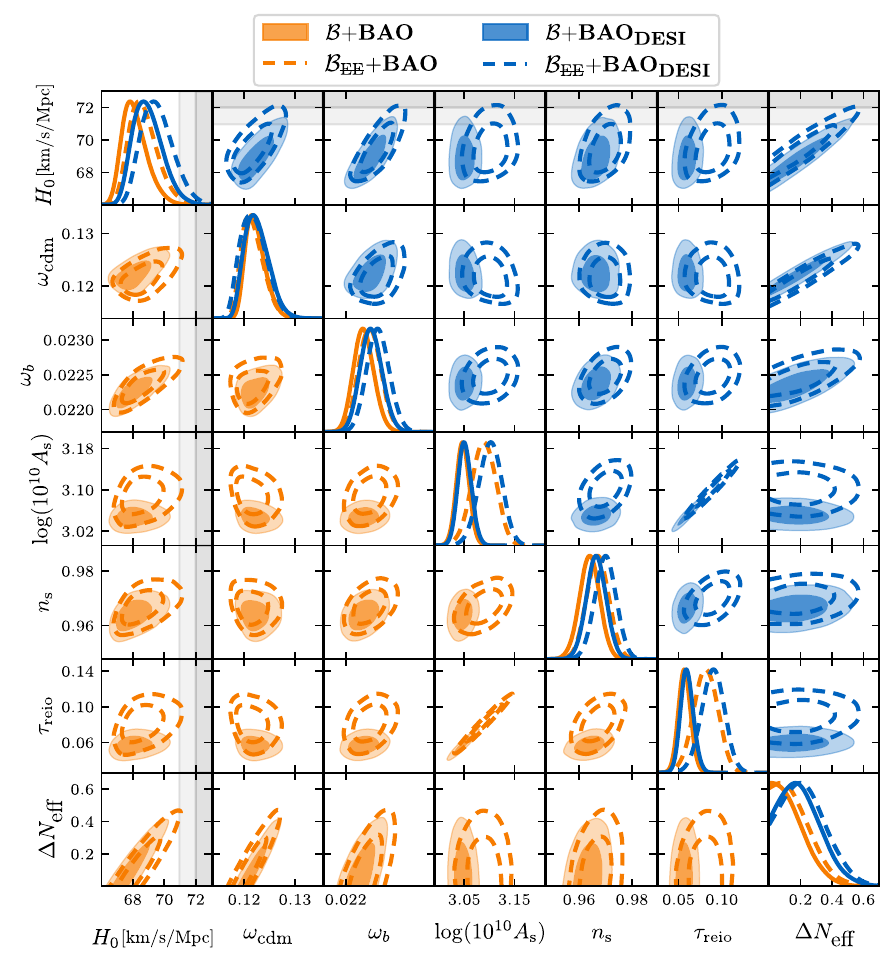}
    \caption{Posterior distributions of the cosmological parameters in the SIDR model along with its new physics parameter, $\DNeff$. The results from the \bao~(\desi) dataset are shown in the lower-left (upper-right) part of the plot as orange (blue) contours. Solid (dashed) contours represent the results obtained from the \baseline~(\baselinewoee) dataset. The gray shaded area shows the $1\sigma$ (dark gray) and $2\sigma$ (light gray) range of $H_0$ from the SH0ES measurement.}
    \label{fig:SIDR_bao_desi}
\end{figure}

\begin{figure}[]
    \centering
    \includegraphics[scale=1.0]{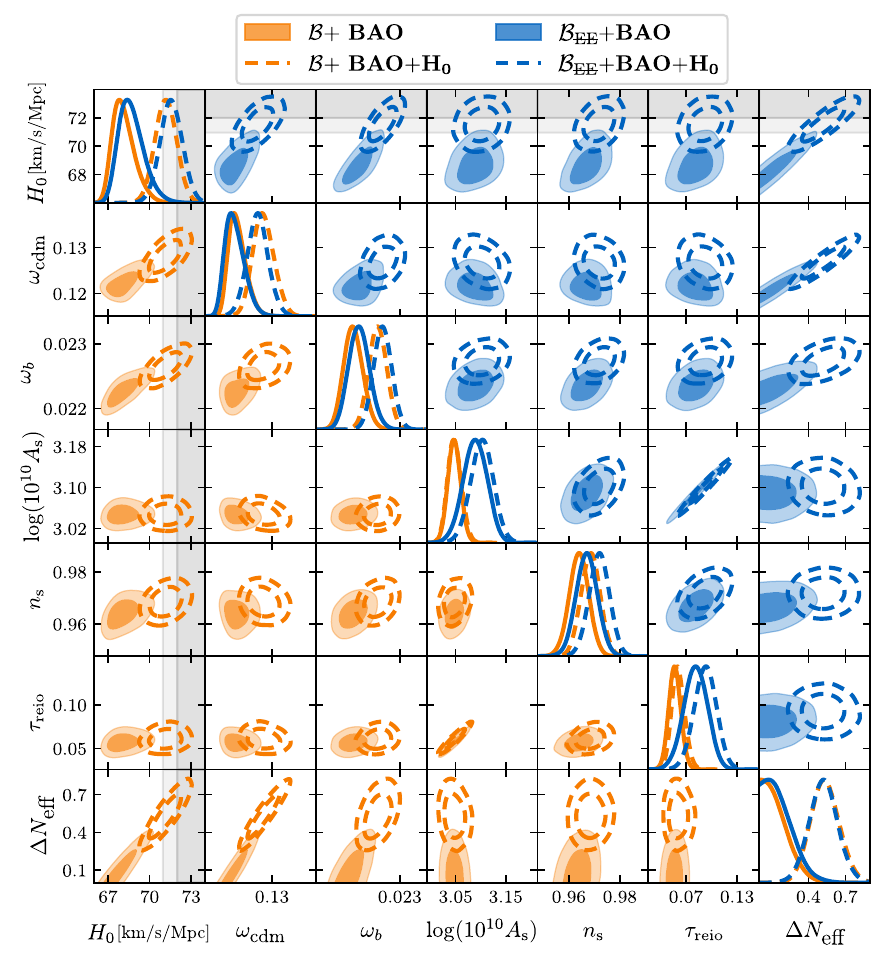}
    \caption{Posterior distributions of the cosmological parameters in the SIDR model along with its new physics parameter, $\DNeff$. The orange (blue) contours show the \baseline+\bao~(\baselinewoee+\bao) dataset. In dashed contours, the \shoes~dataset is included, pulling the posteriors to the high $H_0$ region. The gray shaded area shows the $1\sigma$ (dark gray) and $2\sigma$ (light gray) range of $H_0$ from the SH0ES measurement.
    }
    \label{fig:SIDR_bao_shoes}
\end{figure}

\begin{figure}[H]
    \centering
    \includegraphics[scale=1.0]{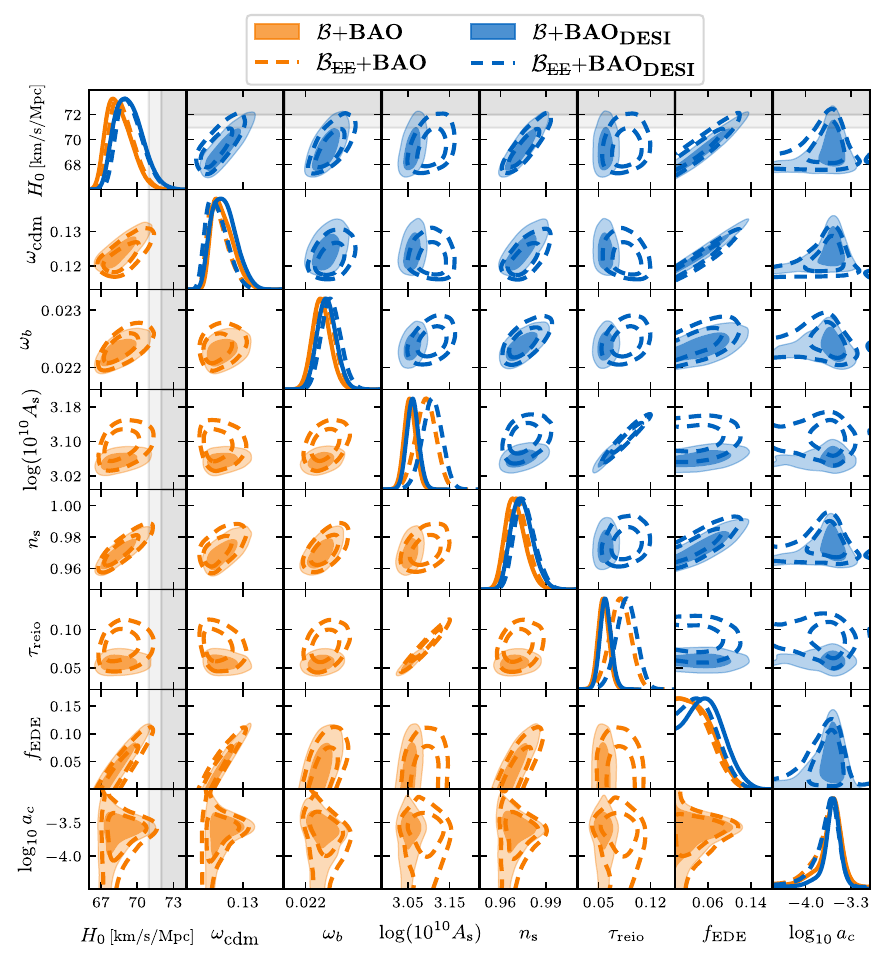}
    \caption{Posterior distributions of the cosmological parameters in the EDE model, along with its new physics parameters, $\left \{\fede, a_c \right\}$. The results from the \bao~(\desi) dataset are shown in the lower-left (upper-right) part of the plot as orange (blue) contours. Solid (dashed) contours represent the results obtained from the \baseline~(\baselinewoee) dataset. The gray shaded area shows the $1\sigma$ (dark gray) and $2\sigma$ (light gray) range of $H_0$ from the SH0ES measurement. }
    \label{fig:EDE_bao_desi}
\end{figure}

\begin{figure}[]
    \centering
    \includegraphics[scale=1.0]{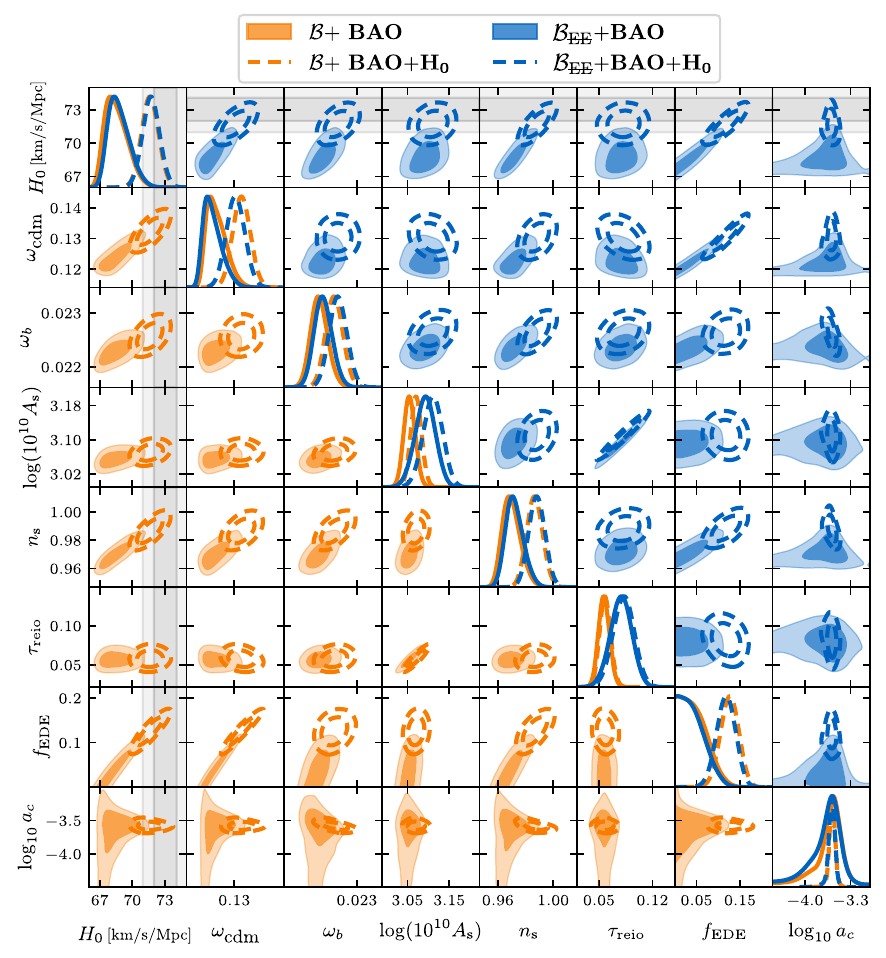}
    \caption{Posterior distributions of the cosmological parameters in the EDE model, along with its new physics parameters, $\left \{\fede, a_c \right\}$. The orange (blue) contours show the \baseline+\bao~(\baselinewoee+\bao) dataset. In dashed contours, the \shoes~dataset is included, pulling the posteriors to the high $H_0$ region. The gray shaded area shows the $1\sigma$ (dark gray) and $2\sigma$ (light gray) range of $H_0$ from the SH0ES measurement.
    }
    \label{fig:EDE_bao_shoes}
    \end{figure}

\cref{app:lcdm} covers the $\Lambda$CDM model, with posterior distributions plotted in \cref{fig:LCDM_bao_desi,fig:LCDM_bao_shoes} and marginalized statistics in \cref{tab:LCDM_bao_desi,tab:LCDM_bao_shoes}. In addition, we obtain best-fit parameters for each analysis via likelihood maximization, utilizing once more the \texttt{Cobaya} sampler. \cref{tab: LCDM_bao_desi_bestfit,tab: LCDM_bao_shoes_bestfit} provide the best-fit parameter values while \cref{tab:LCDM_bao_desi_chi2,tab: LCDM_bao_shoes_chi2} provide the minimized effective $\chi^2$ for each likelihood. Shorthand notations for the likelihood names in \texttt{Cobaya} are given in \cref{tab:likelihoods}.
Subsequently, \cref{app: DR} covers the DR models, with FSDR model depicted in \cref{fig:fsdr_bao_desi,fig:fsdr_bao_shoes} and \cref{tab:fsdr_bao_desi,tab:fsdr_bao_shoes,tab:fsdr_bao_desi_bestfit,tab:fsdr_bao_shoes_bestfit,tab:fsdr_bao_desi_chi2,tab:fsdr_bao_shoes_chi2}, while SIDR results are shown in \cref{fig:SIDR_bao_desi,fig:SIDR_bao_shoes} and \cref{tab:sidr_bao_desi,tab:sidr_bao_shoes,tab:sidr_bao_desi_bestfit,tab:sidr_bao_shoes_bestfit,tab:sidr_bao_desi_chi2,tab:sidr_bao_shoes_chi2}.
Finally, \cref{app:ede} covers the EDE model via \cref{fig:EDE_bao_desi,fig:EDE_bao_shoes} and \cref{tab:ede_bao_desi,tab:ede_bao_shoes,tab:ede_bao_desi_bestfit,tab:ede_bao_shoes_bestfit,tab:ede_bao_desi_chi2,tab:ede_bao_shoes_chi2}.

Let us comment briefly on the impact of our analysis on the value of the matter clustering parameter $S_8 = \sigma_8 \sqrt{\Omega_m/0.3}$. There is a mild tension between measurements of this parameter by weak lensing surveys and the inferred value from fitting $\Lambda$CDM to CMB data. The status of this tension is unclear, since, for example, the combination of the KiDS-1000 and DES Y3 datasets yields a mild $1.7\sigma$ disagreement with CMB observations \cite{Kilo-DegreeSurvey:2023gfr}. Yet, it is important to assess this tension in scenarios with increasing $\taureio$ and increasing $H_0$ as these parameters are known to make the $S_8$ tension worse.\footnote{We thank David Spergel for pointing out the increase of $S_8$ as a result of increasing $\taureio$.} In our analysis, we observe that $S_8$ is slightly increased with the removal of low-$\ell$ EE data. However, the tension remains $\lesssim 2.5\sigma$ in each analysis we presented for $\Lambda$CDM and the DR models. For EDE, which is known to increase $S_8$, the tension still remains below $3\sigma$ without the low-$\ell$ EE data. 

\pagebreak

\subsection{$\Lambda$CDM}\label{app:lcdm}
\FloatBarrier

\label{app:moreplots}

\begin{table}[]
\begin{center}
\begin{tabular} {| l | c| c| c| c|}
\hline
 \multicolumn{1}{|c|}{\bf } &  \multicolumn{1}{|c|}{\baseline +\bao} &  \multicolumn{1}{|c|}{\bf \baselinewoee
+ \bao} &  \multicolumn{1}{|c|}{\baseline +\desi} &  \multicolumn{1}{|c|}{\baselinewoee +\desi}\\
\hline
{$\log(10^{10} A_\mathrm{s})$} & $3.049\pm 0.013            $ & $3.089\pm 0.024            $ & 
$3.054^{+0.012}_{-0.014}   $ & $3.104\pm 0.023            $\\

{$n_\mathrm{s}   $} & $0.9631\pm 0.0036          $ & $0.9660\pm 0.0040          $ & $0.9655\pm 0.0036     
$ & $0.9688\pm 0.0038          $\\

{$\tau_\mathrm{reio}$} & $0.0565^{+0.0066}_{-0.0074}$ & $0.080\pm 0.014            $ & 
$0.0596^{+0.0066}_{-0.0078}$ & $0.088\pm 0.013            $\\

$H_0         \,[\mathrm{km}/\mathrm{s}/\mathrm{Mpc}]               $ & $67.18\pm 0.38             $ & $67.58\pm 0.44             $ & $67.60\pm 0.37        
$ & $68.05\pm 0.40             $\\

$\omegacdm$             & $0.11983\pm 0.00085        $ & $0.11894\pm 0.00097        $ & $0.11888\pm 0.00082     
$ & $0.11790\pm 0.00089        $\\

$\omega_{b}$               & $0.02216\pm 0.00013        $ & $0.02223\pm 0.00014        $ & $0.02223\pm 0.00013     
$ & $0.02232\pm 0.00013        $\\
\hline
$H_0$ Tension& $5.3\sigma $ & $4.8\sigma $ & $4.9\sigma $ & $4.5\sigma $ \\
\hline
\end{tabular}
\end{center}
\caption{The posterior central values and corresponding 68\% C.L. intervals for the cosmological parameters in the $\Lambda$CDM model, along with the $H_0$ tension values from \cref{eq:H0tension}.
The results are presented for \baseline~and \baselinewoee~in combination with the \bao~and the \desi~datasets, as shown in \cref{fig:LCDM_bao_desi}.}
\label{tab:LCDM_bao_desi}
\end{table}


\begin{table}[]
\begin{center}
\begin{tabular} {| l | c| c|}
\hline
 \multicolumn{1}{|c|}{\bf } &  \multicolumn{1}{|c|}{\bf \baseline +\bao +\shoes} &  \multicolumn{1}{|c|}{\bf \baselinewoee +\bao +\shoes}\\
\hline
{$\log(10^{10} A_\mathrm{s})$} & $3.059\pm 0.014            $ & $3.118\pm 0.023            $\\
{$n_\mathrm{s}   $} & $0.9677\pm 0.0035          $ & $0.9715\pm 0.0038          $\\
{$\tau_\mathrm{reio}$} & $0.0626^{+0.0071}_{-0.0082}$ & $0.097\pm 0.013            $\\
$H_0    \,[\mathrm{km}/\mathrm{s}/\mathrm{Mpc}]                    $ & $68.04\pm 0.36             $ & $68.53\pm 0.38             $\\
$\omegacdm$             & $0.11805\pm 0.00080        $ & $0.11695\pm 0.00086        $\\
$\omega_{b}$               & $0.02235\pm 0.00012        $ & $0.02243\pm 0.00012        $\\
\hline
$H_0$ Tension& $4.6\sigma $ & $4.1\sigma $ \\
\hline
\end{tabular}
\end{center}
\caption{The posterior central values and corresponding 68\% C.L. intervals for the cosmological parameters in the $\Lambda$CDM model, along with the $H_0$ tension values from \cref{eq:H0tension}. The results are presented for \baseline~and \baselinewoee~in combination with \bao+\shoes~dataset, as shown in \cref{fig:LCDM_bao_shoes}.} 
\label{tab:LCDM_bao_shoes}
\end{table}

\begin{table}[]
\begin{center}
\begin{tabular} {| l | c| c| c| c|}
\hline
 \multicolumn{1}{|c|}{\bf } &  \multicolumn{1}{|c|}{\baseline +\bao} &  \multicolumn{1}{|c|}{\bf \baselinewoee
+ \bao} &  \multicolumn{1}{|c|}{\baseline +\desi} &  \multicolumn{1}{|c|}{\baselinewoee +\desi}\\
\hline
{$\log(10^{10} A_\mathrm{s})$} & $3.041$ & $3.098$ & $3.056$ & $3.110$ \\
{$n_\mathrm{s}   $} & $0.9638$ & $0.9643$ & $0.9643$ & $0.9684$ \\
{$\tau_\mathrm{reio}$} & $0.0505$ & $0.0851$ & $0.0551$ & $0.0938$ \\
$H_0         \,[\mathrm{km}/\mathrm{s}/\mathrm{Mpc}]               $ & $67.31$ & $67.57$ & $67.82$ & $68.07$ \\
$\omegacdm$             & $0.11959$ & $0.11905$ & $0.11824$ & $0.11800$ \\
$\omega_{b}$               & $0.02218$ & $0.02225$ & $0.02224$ & $0.02236$ \\
\hline
\end{tabular}
\end{center}
\caption{The best-fit values for the cosmological parameters in the $\Lambda$CDM model. The results are presented for \baseline~and \baselinewoee~in combination with the \bao~and the \desi~datasets.}
\label{tab: LCDM_bao_desi_bestfit}
\end{table}

\begin{table}[]
\begin{center}
\begin{tabular} {| l | c| c|}
\hline
 \multicolumn{1}{|c|}{\bf } &  \multicolumn{1}{|c|}{\bf \baseline +\bao +\shoes} &  \multicolumn{1}{|c|}{\bf \baselinewoee +\bao +\shoes}\\
\hline
{$\log(10^{10} A_\mathrm{s})$} & $3.061$ & $3.115$\\
{$n_\mathrm{s}   $} & $0.9693$ & $0.9716$\\
{$\tau_\mathrm{reio}$} & $0.0576$ & $0.0953$\\
$H_0    \,[\mathrm{km}/\mathrm{s}/\mathrm{Mpc}]$ & $68.29$ & $68.45$\\
$\omegacdm$ & $0.11754$ & $0.11715$\\
$\omega_{b}$   & $0.02235$ & $0.02243$\\
\hline
\end{tabular}
\end{center}
\caption{The best-fit values for the cosmological parameters in the $\Lambda$CDM model. The results are presented for \baseline~and \baselinewoee~in combination with \bao+\shoes~dataset.}
\label{tab: LCDM_bao_shoes_bestfit}
\end{table}

\begin{table}[]
\begin{center}
\begin{tabular} {| l | c| c| c| c|}
\hline
 \multicolumn{1}{|c|}{\bf } &  \multicolumn{1}{|c|}{\baseline +\bao} &  \multicolumn{1}{|c|}{\baselinewoee +\bao} &  \multicolumn{1}{|c|}{\baseline +\desi} &  \multicolumn{1}{|c|}{\baselinewoee +\desi}\\
\hline
\PTT & 23.2 & 24.7 & 23.3 & 24.2 \\
\PEE & 395.7 & \na & 396.2 & \na \\
 \PTTTEEE & 10543.7 & 10541.6 & 10544.2 & 10542.4 \\ 
 \act & 22.9 & 22.0 & 22.7 & 21.9 \\
 \sixdf & 0.07 & 0.03 & \na & \na \\ 
 \mgs & 1.0 & 1.2 & \na & \na \\
 \consensus & 5.2 & 4.4 & \na & \na \\ 
\pantheon & 1403.8 & 1404.2 & 1404.8 & 1405.2 \\ 
\BAOdesi & \na & \na & 15.8 & 14.8 \\ 
\hline
 $\chi^2_{\mathrm{total}}$ &  12395.7 & 11998.2 & 12407.1 & 12008.7 \\ \hline
\end{tabular}
\end{center}
\caption{The contribution to the effective $\chi^2$ from each likelihood when fitting the $\Lambda$CDM model to \baseline~and \baselinewoee~in combination with the \bao~and the \desi~datasets. Shorthand notations for the likelihoods are given in \cref{tab:likelihoods}.}
\label{tab:LCDM_bao_desi_chi2}
\end{table}

\begin{table}[]
\begin{center}
\begin{tabular} {| l | c| c|}
\hline
 \multicolumn{1}{|c|}{\bf } &  \multicolumn{1}{|c|}{\bf \baseline +\bao +\shoes} &  \multicolumn{1}{|c|}{\bf \baselinewoee +\bao +\shoes}\\
\hline
\PTT & 22.4 & 23.6 \\ 
        \PEE & 396.6 & \na \\ 
        \PTTTEEE & 10547.3 & 10544.5 \\ 
        \act & 22.6 & 22.1 \\ 
        \sixdf & 0.01 & 0.02 \\ 
        \mgs & 2.0 & 2.2 \\ 
        \consensus & 3.4 & 3.4 \\ 
        \pantheonshoes & 1486.7 & 1485.4 \\ \hline
$\chi^2_{\mathrm{total}}$  & 12481.1 & 12081.2 \\ 
\hline
\end{tabular}
\end{center}
\caption{The contribution to the effective $\chi^2$ from each likelihood when fitting the $\Lambda$CDM model to \baseline~and \baselinewoee~in combination with the \shoes~dataset. Shorthand notations for the likelihoods are given in \cref{tab:likelihoods}.}
\label{tab: LCDM_bao_shoes_chi2}
\end{table}

\FloatBarrier

\subsection{Dark Radiation}
\label{app: DR}

\subsubsection{FSDR}

\begin{table}[]
\begin{center}
\begin{tabular} {| l | c| c| c| c|}
\hline
 \multicolumn{1}{|c|}{\bf } &  \multicolumn{1}{|c|}{\baseline +\bao} &  \multicolumn{1}{|c|}{\baselinewoee +\bao} &  \multicolumn{1}{|c|}{\baseline +\desi} &  \multicolumn{1}{|c|}{\baselinewoee +\desi}\\
\hline
{$\log(10^{10} A_\mathrm{s})$} & $3.055\pm 0.014            $ & $3.092\pm 0.024            $ & $3.061^{+0.013}_{-0.015}   $ & $3.107\pm 0.023            $\\
{$n_\mathrm{s}   $} & $0.9676^{+0.0044}_{-0.0053}$ & $0.9701^{+0.0047}_{-0.0054}$ & $0.9712^{+0.0048}_{-0.0057}$ & $0.9738^{+0.0049}_{-0.0056}$\\
{$\tau_\mathrm{reio}$} & $0.0566\pm 0.0071          $ & $0.078\pm 0.014            $ & $0.0591^{+0.0067}_{-0.0078}$ & $0.086\pm 0.013            $\\
$H_0          \,[\mathrm{km}/\mathrm{s}/\mathrm{Mpc}]              $ & $68.13^{+0.58}_{-0.92}     $ & $68.45^{+0.60}_{-0.91}     $ & $68.80^{+0.69}_{-1.0}      $ & $69.12^{+0.65}_{-1.0}      $\\
$\omegacdm$             & $0.1223^{+0.0014}_{-0.0023}$ & $0.1214^{+0.0015}_{-0.0024}$ & $0.1222^{+0.0018}_{-0.0026}$ & $0.1210^{+0.0018}_{-0.0028}$\\
$\omega_{b}$               & $0.02225\pm 0.00014        $ & $0.02231\pm 0.00014        $ & $0.02233\pm 0.00015        $ & $0.02241\pm 0.00015        $\\
$\Delta N_{\mbox{eff}}$    & $< 0.376                   $ & $< 0.370                   $ & $0.206^{+0.064}_{-0.19}    $ & $< 0.436                   $\\
\hline
$H_0$ Tension& $4.1\sigma $ & $3.8\sigma $ & $3.4\sigma $ & $3.2\sigma $ \\
\hline
\end{tabular}
\end{center}
\caption{The posterior central values and corresponding 68\% C.L. intervals for the cosmological parameters in the FSDR model, along with its new physics parameter $\DNeff$ and $H_0$ tension values from~\cref{eq:H0tension}. The results are presented for \baseline~and \baselinewoee~in combination with the \bao~and the \desi~datasets, as shown in \cref{fig:fsdr_bao_desi}. For $\DNeff$, the 95\% C.L. upper bounds are provided only when a credible 68\% interval away from zero is not obtained.}
\label{tab:fsdr_bao_desi}
\end{table}

\begin{table}[]
\begin{center}

\begin{tabular} {| l | c| c|}
\hline
 \multicolumn{1}{|c|}{\bf } &  \multicolumn{1}{|c|}{\bf \baseline +\bao +\shoes} &  \multicolumn{1}{|c|}{\bf \baselinewoee +\bao +\shoes}\\
\hline
{$\log(10^{10} A_\mathrm{s})$} & $3.073\pm 0.014            $ & $3.117\pm 0.023            $\\
{$n_\mathrm{s}   $} & $0.9818\pm 0.0048          $ & $0.9836\pm 0.0049          $\\
{$\tau_\mathrm{reio}$} & $0.0586^{+0.0067}_{-0.0077}$ & $0.085\pm 0.014            $\\
$H_0     \,[\mathrm{km}/\mathrm{s}/\mathrm{Mpc}]                   $ & $71.15\pm 0.79             $ & $71.26\pm 0.78             $\\
$\omegacdm$             & $0.1287\pm 0.0026          $ & $0.1268^{+0.0024}_{-0.0027}$\\
$\omega_{b}$               & $0.02257\pm 0.00013        $ & $0.02262\pm 0.00013        $\\
$\Delta N_{\mbox{eff}}$    & $0.61\pm 0.14              $ & $0.55\pm 0.14              $\\
\hline
$H_0$ Tension& $1.5\sigma $ & $1.4\sigma $ \\
\hline
\end{tabular}
    
\end{center}

\caption{The posterior central values and corresponding 68\% C.L. intervals for the cosmological parameters in the FSDR model, along with its new physics parameter $\DNeff$ and $H_0$ tension values from \cref{eq:H0tension}. The results are presented for \baseline~and \baselinewoee~in combination with \bao+\shoes~dataset, as shown in \cref{fig:fsdr_bao_shoes}.}
\label{tab:fsdr_bao_shoes}
\end{table}

\begin{table}[]
\begin{center}
\begin{tabular} {| l | c| c| c| c|}
\hline
 \multicolumn{1}{|c|}{\bf } &  \multicolumn{1}{|c|}{\baseline +\bao} &  \multicolumn{1}{|c|}{\baselinewoee +\bao} &  \multicolumn{1}{|c|}{\baseline +\desi} &  \multicolumn{1}{|c|}{\baselinewoee +\desi}\\
\hline
{$\log(10^{10} A_\mathrm{s})$} & $3.057$ & $3.103$ & $3.046$ & $3.114$\\
{$n_\mathrm{s}$} & $0.9673$ & $0.9673$ & $0.9673$ & $0.9740$\\
{$\tau_\mathrm{reio}$} & $0.0540$ & $0.0875$ & $0.0504$ & $0.0885$\\
$H_0 \,[\mathrm{km}/\mathrm{s}/\mathrm{Mpc}]$ & 68.34 & 67.85 & 68.22 & 69.33 \\
$\omegacdm$ & $0.12208$ & $0.11930$ & $0.12080$ & $0.12140$\\
$\omega_{b}$ & $0.02223$ & $0.02227$ & $0.02223$ & $0.02241$\\
$\Delta N_{\mbox{eff}}$ & $0.17$ & $0.03$ & $0.11$ & $0.23$\\
\hline
$\chi^2-\chi_{\Lambda \rm{CDM}}^2$ & $-0.6$ & $+0.0$ & $+0.3$ & $+0.3$ \\
\hline
\end{tabular}
\end{center}
\caption{The best-fit values for the cosmological parameters in the FSDR model, along with its new physics parameter $\DNeff$. The results are presented for \baseline~and \baselinewoee~in combination with the \bao~and the \desi~datasets. The $\chi^2$ differences with respect to the corresponding $\Lambda$CDM cases are also presented.}
\label{tab:fsdr_bao_desi_bestfit}
\end{table}

\begin{table}[]
\begin{center}
\begin{tabular} {| l | c| c|}
\hline
 \multicolumn{1}{|c|}{\bf Parameter} &  \multicolumn{1}{|c|}{\bf \baseline +\bao +\shoes} &  \multicolumn{1}{|c|}{\bf \baselinewoee +\bao +\shoes}\\
\hline
{$\log(10^{10} A_\mathrm{s})$} & $3.077$ & $3.115$\\
{$n_\mathrm{s}$} & $0.9804$ & $0.9817$\\
{$\tau_\mathrm{reio}$} & $0.0566$ & $0.0823$\\
$H_0 \,[\mathrm{km}/\mathrm{s}/\mathrm{Mpc}]$ & $71.28$ & $71.14$\\
$\omegacdm$ & $0.12873$ & $0.12642$\\
$\omega_{b}$ & $0.02256$ & $0.02258$\\
$\Delta N_{\text{eff}}$ & $0.63$ & $0.53$\\
\hline
$\chi^2-\chi_{\Lambda \rm{CDM}}^2$ & $-21.1$ & $-18.8$ \\ \hline
\end{tabular}
\end{center}
\caption{The best-fit values for the cosmological parameters in the FSDR model, along with its new physics parameter $\DNeff$. The results are presented for \baseline~and \baselinewoee~in combination with \bao+\shoes~dataset. The $\chi^2$ differences with respect to the corresponding $\Lambda$CDM cases are also presented.}
\label{tab:fsdr_bao_shoes_bestfit}
\end{table}

\begin{table}[]
\begin{center}
\begin{tabular} {| l | c| c| c| c|}
\hline
 \multicolumn{1}{|c|}{\bf } &  \multicolumn{1}{|c|}{\baseline +\bao} &  \multicolumn{1}{|c|}{\baselinewoee +\bao} &  \multicolumn{1}{|c|}{\baseline +\desi} &  \multicolumn{1}{|c|}{\baselinewoee +\desi}\\
\hline
\PTT & 23.0 & 24.2 & 22.7 & 23.2 \\ 
        \PEE & 396.1 & \na & 395.8 & \na \\ 
        \PTTTEEE & 10543.9 & 10542.2 & 10544.3 & 10543.9 \\ 
        \act & 22.1 & 21.9 & 23.2 & 22.1 \\ 
        \sixdf & 0.03 & 0.01 & \na & \na \\ 
        \mgs & 1.3 & 1.4 & \na & \na \\ 
        \consensus & 4.4 & 4.0 & \na & \na \\ 
        \pantheon & 1404.3 & 1404.5 & 1404.6 & 1405.9 \\ 
        \BAOdesi & \na & \na & 16.8 & 13.9 \\ \hline
$\chi^2_{\mathrm{total}}$ & 12395.1 & 11998.1 & 12407.4 & 12009.0 \\ \hline
\end{tabular}
\end{center}
\caption{The contribution to the effective $\chi^2$ from each likelihood when fitting the FSDR model to \baseline~and \baselinewoee~in combination with the \bao~and the \desi~datasets. Shorthand notations for the likelihoods are given in \cref{tab:likelihoods}.}
\label{tab:fsdr_bao_desi_chi2}
\end{table}

\begin{table}[]
\begin{center}
\begin{tabular} {| l | c| c| c| c|}
\hline
 \multicolumn{1}{|c|}{\bf } &  \multicolumn{1}{|c|}{\bf \baseline +\bao +\shoes} &  \multicolumn{1}{|c|}{\bf \baselinewoee +\bao +\shoes}\\
\hline
\PTT & 21.6 & 22.2 \\ 
        \PEE & 396.5 & \na \\ 
        \PTTTEEE & 10552.1 & 10548.7 \\ 
        \act & 22.2 & 22.0 \\ 
        \sixdf & 0.01 & 0.04 \\ 
        \mgs & 2.1 & 2.4 \\ 
        \consensus & 3.5 & 3.6 \\ 
        \pantheonshoes & 1461.8 & 1463.5 \\ \hline
$\chi^2_{\mathrm{total}}$ & 12460.0 & 12062.4 \\ \hline
\end{tabular}
\end{center}
\caption{The contribution to the effective $\chi^2$ from each likelihood when fitting the FSDR model to \baseline~and \baselinewoee~in combination with the \shoes~dataset. Shorthand notations for the likelihoods are given in \cref{tab:likelihoods}.}
\label{tab:fsdr_bao_shoes_chi2}
\end{table}

\FloatBarrier

\subsubsection{SIDR}\label{app:sidr}

\begin{table}[]
\begin{center}
\begin{tabular} {| l | c| c| c| c|}
\hline
 \multicolumn{1}{|c|}{\bf } &  \multicolumn{1}{|c|}{\baseline +\bao} &  \multicolumn{1}{|c|}{\baselinewoee +\bao} &  \multicolumn{1}{|c|}{\baseline +\desi} &  \multicolumn{1}{|c|}{\baselinewoee +\desi}\\
\hline
{$\log(10^{10} A_\mathrm{s})$} & $3.047\pm 0.013            $ & $3.089\pm 0.024            $ & $3.051\pm 0.014            $ & $3.102\pm 0.023            $\\
{$n_\mathrm{s}   $} & $0.9640\pm 0.0038          $ & $0.9670\pm 0.0042          $ & $0.9665\pm 0.0037          $ & $0.9700\pm 0.0040          $\\
{$\tau_\mathrm{reio}$} & $0.0570\pm 0.0071          $ & $0.081\pm 0.014            $ & $0.0599^{+0.0068}_{-0.0078}$ & $0.090\pm 0.013            $\\
$H_0       \,[\mathrm{km}/\mathrm{s}/\mathrm{Mpc}]                 $ & $68.14^{+0.59}_{-0.94}     $ & $68.69^{+0.69}_{-1.1}      $ & $68.95^{+0.74}_{-1.1}      $ & $69.55^{+0.82}_{-1.2}      $\\
$\omegacdm$             & $0.1223^{+0.0014}_{-0.0023}$ & $0.1217^{+0.0016}_{-0.0025}$ & $0.1225^{+0.0019}_{-0.0028}$ & $0.1219^{+0.0021}_{-0.0029}$\\
$\omega_{b}$               & $0.02227^{+0.00014}_{-0.00016}$ & $0.02236\pm 0.00016        $ & $0.02238\pm 0.00016        $ & $0.02248\pm 0.00017        $\\
$\Delta N_{\mbox{eff}}$    & $< 0.340                   $ & $< 0.380                   $ & $0.206^{+0.078}_{-0.18}    $ & $0.228^{+0.096}_{-0.18}    $\\
\hline
$H_0$ Tension& $4.1\sigma $ & $3.5\sigma $ & $3.2\sigma $ & $2.6\sigma $ \\
\hline
\end{tabular}
\end{center}
\caption{The posterior central values and corresponding 68\% C.L. intervals for the cosmological parameters in the SIDR model, along with its new physics parameter $\DNeff$ and $H_0$ tension values from \cref{eq:H0tension}. The results are presented for \baseline~and \baselinewoee~in combination with the \bao~and the \desi~datasets, as shown in \cref{fig:SIDR_bao_desi}. For $\DNeff$, the 95\% C.L. upper bounds are provided only when a credible 68\% interval away from zero is not obtained.}
\label{tab:sidr_bao_desi}
\end{table}


\begin{table}[]
\begin{center}

\begin{tabular} {| l | c| c|}
\hline
 \multicolumn{1}{|c|}{\bf } &  \multicolumn{1}{|c|}{\bf \baseline +\bao +\shoes} &  \multicolumn{1}{|c|}{\bf \baselinewoee +\bao +\shoes}\\
\hline
{$\log(10^{10} A_\mathrm{s})$} & $3.048^{+0.013}_{-0.015}   $ & $3.102\pm 0.023            $\\
{$n_\mathrm{s}   $} & $0.9685\pm 0.0037          $ & $0.9719\pm 0.0040          $\\
{$\tau_\mathrm{reio}$} & $0.0611^{+0.0069}_{-0.0080}$ & $0.093\pm 0.013            $\\
$H_0       \,[\mathrm{km}/\mathrm{s}/\mathrm{Mpc}]                 $ & $71.17\pm 0.77             $ & $71.57\pm 0.80             $\\
$\omegacdm$             & $0.1281^{+0.0022}_{-0.0025}$ & $0.1268\pm 0.0024          $\\
$\omega_{b}$               & $0.02265\pm 0.00014        $ & $0.02273\pm 0.00014        $\\
$\Delta N_{\mbox{eff}}$    & $0.53\pm 0.12              $ & $0.52\pm 0.12              $\\
\hline
$H_0$ Tension& $1.5\sigma $ & $1.1\sigma $ \\
\hline
\end{tabular}
    
\end{center}
\caption{The posterior central values and corresponding 68\% C.L. intervals for the cosmological parameters in the SIDR model, along with its new physics parameter $\DNeff$ and $H_0$ tension values from \cref{eq:H0tension}. The results are presented for \baseline~and \baselinewoee~in combination with \bao+\shoes~dataset, as shown in \cref{fig:SIDR_bao_shoes}.}
\label{tab:sidr_bao_shoes}
\end{table}

\begin{table}[]
\begin{center}
\begin{tabular} {| l | c| c| c| c|}
\hline
 \multicolumn{1}{|c|}{\bf } &  \multicolumn{1}{|c|}{\baseline +\bao} &  \multicolumn{1}{|c|}{\baselinewoee +\bao} &  \multicolumn{1}{|c|}{\baseline +\desi} &  \multicolumn{1}{|c|}{\baselinewoee +\desi}\\
\hline
{$\log(10^{10} A_\mathrm{s})$} & $3.055$ & $3.093$ & $3.043$ & $3.101$\\
{$n_\mathrm{s}   $} & $0.9639$ & $0.9675$ & $0.9670$ & $0.9705$\\
{$\tau_\mathrm{reio}$} & $0.0577$ & $0.0868$ & $0.0476$ & $0.0893$\\
$H_0          \,[\mathrm{km}/\mathrm{s}/\mathrm{Mpc}]$              & 67.97 & 68.98 & 69.38 & 70.09 \\
$\omegacdm$             & $0.12222$ & $0.12230$ & $0.12222$ & $0.12370$\\
$\omega_{b}$               & $0.02222$ & $0.02241$ & $0.02242$ & $0.02251$\\
$\Delta N_{\mbox{eff}}$    & $0.11$ & $0.20$ & $0.23$ & $0.33$\\
\hline
$\chi^2-\chi_{\Lambda \rm{CDM}}^2$ & $-1.0$ & $+0.1$ & $-2.9$ & $-1.8$ \\ \hline
\end{tabular}
\end{center}
\caption{The best-fit values for the cosmological parameters in the SIDR model, along with its new physics parameter $\DNeff$.  The results are presented for \baseline~and \baselinewoee~in combination with the \bao~and the \desi~datasets. The $\chi^2$ differences with respect to the corresponding $\Lambda$CDM cases are also presented.}
\label{tab:sidr_bao_desi_bestfit}
\end{table}

\begin{table}[]
\begin{center}
\begin{tabular} {| l | c| c|}
\hline
 \multicolumn{1}{|c|}{\bf } &  \multicolumn{1}{|c|}{\bf \baseline +\bao +\shoes} &  \multicolumn{1}{|c|}{\bf \baselinewoee +\bao +\shoes}\\
\hline
{$\log(10^{10} A_\mathrm{s})$} & $3.040$ & $3.093$\\
{$n_\mathrm{s}   $} & $0.9702$ & $0.9701$\\
{$\tau_\mathrm{reio}$} & $0.0534$ & $0.0861$\\
$H_0       \,[\mathrm{km}/\mathrm{s}/\mathrm{Mpc}]$ & $71.75$ & $71.83$\\
$\omegacdm$ & $0.12916$ & $0.12841$\\
$\omega_{b}$ & $0.02268$ & $0.02275$\\
$\Delta N_{\mbox{eff}}$ & $0.61$ & $0.59$\\
\hline
$\chi^2-\chi_{\Lambda \rm{CDM}}^2$ & $-24.1$ & $-23.1$ \\ \hline
\end{tabular}
\end{center}
\caption{The best-fit values for the cosmological parameters in the SIDR model, along with its new physics parameter $\DNeff$. The results are presented for \baseline~and \baselinewoee~in combination with \bao+\shoes~dataset. The $\chi^2$ differences with respect to the corresponding $\Lambda$CDM cases are also presented.}
\label{tab:sidr_bao_shoes_bestfit}
\end{table}

\begin{table}[]
\begin{center}
\begin{tabular} {| l | c| c| c| c|}
\hline
 \multicolumn{1}{|c|}{\bf } &  \multicolumn{1}{|c|}{\baseline +\bao} &  \multicolumn{1}{|c|}{\baselinewoee +\bao} &  \multicolumn{1}{|c|}{\baseline +\desi} &  \multicolumn{1}{|c|}{\baselinewoee +\desi}\\
\hline
\PTT & 23.4 & 23.9 & 22.4 & 23.2 \\ 
\PEE & 397.0 & \na & 395.8 & \na \\ 
\PTTTEEE & 10542.2 & 10541.8 & 10543.4 & 10542.0 \\ 
 \act & 22.1 & 22.3 & 22.8 & 22.2 \\ 
 \sixdf & 0.05 & 0.00 & \na & \na \\ 
 \mgs & 1.1 & 1.7 & \na & \na \\ 
 \consensus & 4.8 & 3.6 & \na & \na \\ 
 \pantheon & 1404.0 & 1405.0 & 1405.7 & 1406.3 \\ 
\BAOdesi & \na & \na & 14.0 & 13.3 \\ \hline
$\chi^2_{\mathrm{total}}$ & 12394.7 & 11998.2 & 12404.1 & 12006.8 \\ \hline
\end{tabular}
\end{center}
\caption{The contribution to the effective $\chi^2$ from each likelihood when fitting the SIDR model to \baseline~and \baselinewoee~in combination with the \bao~and the \desi~datasets. Shorthand notations for the likelihoods are given in \cref{tab:likelihoods}.}
\label{tab:sidr_bao_desi_chi2}
\end{table}

\begin{table}[]
\begin{center}
\begin{tabular} {| l | c| c|}
\hline
 \multicolumn{1}{|c|}{\bf } &  \multicolumn{1}{|c|}{\bf \baseline +\bao +\shoes} &  \multicolumn{1}{|c|}{\bf \baselinewoee +\bao +\shoes}\\
\hline
\PTT & 21.9 & 23.0 \\ 
\PEE & 396.0 & \na \\ 
 \PTTTEEE & 10549.1 & 10545.2 \\ 
 \act & 23.6 & 22.7 \\ 
 \sixdf & 0.05 & 0.08 \\ 
 \mgs & 2.4 & 2.6 \\ 
\consensus & 3.7 & 3.9 \\ 
 \pantheonshoes & 1460.4 & 1460.6 \\ \hline
$\chi^2_{\mathrm{total}}$ & 12457.0 & 12058.1 \\ \hline
\end{tabular}
\end{center}
\caption{The contribution to the effective $\chi^2$ from each likelihood when fitting the SIDR model to \baseline~and \baselinewoee~in combination with the \shoes~dataset. Shorthand notations for the likelihoods are given in \cref{tab:likelihoods}.}
\label{tab:sidr_bao_shoes_chi2}
\end{table}

\FloatBarrier

\subsection{Early Dark Energy }\label{app:ede}

\begin{table}[]

\begin{tabular} {| l | c| c| c| c|}
\hline
 \multicolumn{1}{|c|}{\bf } &  \multicolumn{1}{|c|}{\baseline +\bao} &  \multicolumn{1}{|c|}{\baselinewoee +\bao} &  \multicolumn{1}{|c|}{\baseline +\desi} &  \multicolumn{1}{|c|}{\baselinewoee +\desi}\\
\hline
{$\log(10^{10} A_\mathrm{s})$} & $3.055\pm 0.014            $ & $3.093\pm 0.024            $ & $3.062\pm 0.014            $ & $3.108\pm 0.023            $\\
{$n_\mathrm{s}   $} & $0.9699^{+0.0054}_{-0.0074}$ & $0.9721^{+0.0054}_{-0.0072}$ & $0.9744^{+0.0059}_{-0.0079}$ & $0.9762^{+0.0056}_{-0.0077}$\\
{$H_0   \,[\mathrm{km}/\mathrm{s}/\mathrm{Mpc}]          $} & $68.52^{+0.77}_{-1.2}      $ & $68.78^{+0.74}_{-1.2}      $ & $69.32^{+0.92}_{-1.3}      $ & $69.43^{+0.80}_{-1.2}      $\\
{$\tau_\mathrm{reio}$} & $0.0566\pm 0.0073          $ & $0.079\pm 0.014            $ & $0.0594^{+0.0069}_{-0.0077}$ & $0.087\pm 0.014            $\\
$\omegacdm$             & $0.1240^{+0.0022}_{-0.0039}$ & $0.1228^{+0.0020}_{-0.0036}$ & $0.1244^{+0.0028}_{-0.0043}$ & $0.1224^{+0.0023}_{-0.0039}$\\
$\omega_{b}$               & $0.02229^{+0.00015}_{-0.00017}$ & $0.02236^{+0.00016}_{-0.00018}$ & $0.02238^{+0.00016}_{-0.00018}$ & $0.02246^{+0.00016}_{-0.00019}$\\
$\fede$         & $< 0.0977                  $ & $< 0.0924                   $ & $0.056^{+0.028}_{-0.041}   $ & $0.048^{+0.016}_{-0.045}   $\\
$\log_{10}a_c              $ & $-3.64^{+0.25}_{-0.11}     $ & $-3.68^{+0.27}_{-0.15}     $ & $-3.62^{+0.19}_{-0.081}    $ & $-3.68^{+0.24}_{-0.13}     $\\
\hline
$H_0$ Tension& $3.5\sigma $ & $3.3\sigma $ & $2.7\sigma $ & $2.8\sigma $ \\
\hline
\end{tabular}

\caption{The posterior central values and corresponding 68\% C.L. intervals for the cosmological parameters in the EDE model, along with its new physics parameters $\left\{\fede, a_c \right\}$ and $H_0$ tension values from \cref{eq:H0tension}. The results are presented for \baseline~and \baselinewoee~in combination with the \bao~and the \desi~datasets, as shown in \cref{fig:EDE_bao_desi}. For $\fede$, the 95\% C.L. upper bounds are provided only when a credible 68\% interval away from zero is not obtained.}
\label{tab:ede_bao_desi}
\end{table}



\begin{table}[]
\begin{center}
\begin{tabular} {| l | c| c|}
\hline
 \multicolumn{1}{|c|}{\bf } &  \multicolumn{1}{|c|}{\bf \baseline +\bao +\shoes} &  \multicolumn{1}{|c|}{\bf \baselinewoee +\bao +\shoes}\\
\hline
{$\log(10^{10} A_\mathrm{s})$} & $3.071\pm 0.013            $ & $3.111\pm 0.023            $\\
{$n_\mathrm{s}   $} & $0.9866\pm 0.0058          $ & $0.9882\pm 0.0058          $\\
{$H_0     \,[\mathrm{km}/\mathrm{s}/\mathrm{Mpc}]        $} & $71.64\pm 0.78             $ & $71.73\pm 0.79             $\\
{$\tau_\mathrm{reio}$} & $0.0581\pm 0.0073          $ & $0.082\pm 0.014            $\\
$\omegacdm$             & $0.1324\pm 0.0029          $ & $0.1304\pm 0.0031          $\\
$\omega_{b}$               & $0.02257\pm 0.00016        $ & $0.02264\pm 0.00017        $\\
$f_{\mathrm{EDE}}$         & $0.126\pm 0.021            $ & $0.118\pm 0.022            $\\
$\log_{10}a_c              $ & $-3.580\pm 0.046           $ & $-3.591^{+0.051}_{-0.042}  $\\
\hline
$H_0$ Tension& $1.1\sigma $ & $1.0\sigma $ \\
\hline
\end{tabular}
\end{center}
\caption{The posterior central values and corresponding 68\% C.L. intervals for the cosmological parameters in the EDE model, along with its new physics parameters $\left\{\fede, a_c \right\}$ and $H_0$ tension values from \cref{eq:H0tension}. The results are presented for \baseline~and \baselinewoee~in combination with \bao+\shoes~dataset, as shown in \cref{fig:EDE_bao_shoes}.}
\label{tab:ede_bao_shoes}
\end{table}

\begin{table}[]
\begin{center}
\begin{tabular} {| l | c| c| c| c|}
\hline
 \multicolumn{1}{|c|}{\bf } &  \multicolumn{1}{|c|}{\baseline +\bao} &  \multicolumn{1}{|c|}{\baselinewoee +\bao} &  \multicolumn{1}{|c|}{\baseline +\desi} &  \multicolumn{1}{|c|}{\baselinewoee  +\desi}\\
\hline
{$\log(10^{10} A_\mathrm{s})$} & $3.062$ & $3.086$ & $3.062$ & $3.126$\\
{$n_\mathrm{s}$} & $0.972$ & $0.974$ & $0.976$ & $0.979$\\
{$H_0 \,[\mathrm{km}/\mathrm{s}/\mathrm{Mpc}]$} & 69.21 & 69.34 & 69.69 & 69.63\\
{$\tau_\mathrm{reio}$} & $0.0550$ & $0.0739$ & $0.0559$ & $0.0992$\\
$\omegacdm$ & $0.12453$ & $0.12498$ & $0.12441$ & $0.12387$\\
$\omega_{b}$ & $0.02232$ & $0.02243$ & $0.02233$ & $0.02246$\\
$f_{\mathrm{EDE}}$ & $0.056$ & $0.060$ & $0.061$ & $0.059$\\
$\log_{10}a_c$ & $-3.57$ & $-3.59$ & $-3.52$ & $-3.62$\\
\hline
$\chi^2-\chi_{\Lambda \rm{CDM}}^2$ & $-2.0$ & $-2.0$ & $-1.8$ & $-2.5$ \\
\hline
\end{tabular}
\end{center}
\caption{The best-fit values for the cosmological parameters in the EDE model, along with its new physics parameters $\left\{\fede, a_c \right\}$.  The results are presented for \baseline~and \baselinewoee~in combination with the \bao~and the \desi~datasets. The $\chi^2$ differences with respect to the corresponding $\Lambda$CDM cases are also presented.}
\label{tab:ede_bao_desi_bestfit}
\end{table}

\begin{table}[]
\begin{center}
\begin{tabular} {| l | c| c|}
\hline
 \multicolumn{1}{|c|}{\bf } &  \multicolumn{1}{|c|}{\bf \baseline +\bao +\shoes} &  \multicolumn{1}{|c|}{\bf \baselinewoee +\bao +\shoes}\\
\hline
{$\log(10^{10} A_\mathrm{s})$} & $3.072$ & $3.126$ \\
{$n_\mathrm{s}   $} & $0.985$ & $0.990$ \\
{$H_0     \,[\mathrm{km}/\mathrm{s}/\mathrm{Mpc}]        $} & $71.52$ & $71.58$ \\
{$\tau_\mathrm{reio}$} & $0.0555$ & $0.0963$ \\
$\omegacdm$             & $0.13196$ & $0.12928$ \\
$\omega_{b}$               & $0.02252$ & $0.02272$ \\
$f_{\mathrm{EDE}}$         & $0.123$ & $0.110$ \\
$\log_{10}a_c              $ & $-3.574$ & $-3.610$ \\
\hline
$\chi^2-\chi_{\Lambda \rm{CDM}}^2$ & $-28.1$ & $-27.0$ \\ \hline
\end{tabular}
\end{center}
\caption{The best-fit values for the cosmological parameters in the EDE model, along with its new physics parameters $\left\{\fede, a_c \right\}$. The results are presented for \baseline~and \baselinewoee~in combination with \bao+\shoes~dataset. The $\chi^2$ differences with respect to the corresponding $\Lambda$CDM cases are also presented.}
\label{tab:ede_bao_shoes_bestfit}
\end{table}

\begin{table}[]
\begin{center}
\begin{tabular} {| l | c| c| c| c|}
\hline
 \multicolumn{1}{|c|}{\bf } &  \multicolumn{1}{|c|}{\baseline +\bao} &  \multicolumn{1}{|c|}{\baselinewoee +\bao} &  \multicolumn{1}{|c|}{\baseline +\desi} &  \multicolumn{1}{|c|}{\baselinewoee +\desi}\\
\hline
\PTT & 22.5 & 23.0 & 22.0 & 23.1 \\ 
 \PEE & 396.2 & \na & 396.3 & \na \\ 
  \PTTTEEE & 10543.3 & 10541.9 & 10545.4 & 10541.9 \\ 
 \act & 21.6 & 21.3 & 21.5 & 21.2 \\ 
\sixdf & 0.01 & 0.01 & \na & \na \\ 
 \mgs & 1.5 & 1.4 & \na & \na \\ 
 \consensus & 3.9 & 4.0 & \na & \na \\ 
 \pantheon & 1404.7 & 1404.7 & 1405.4 & 1405.4 \\ 
\BAOdesi & \na & \na & 14.6 & 14.6 \\ \hline
$\chi^2_{\mathrm{total}}$ & 12393.7 & 11996.2 & 12405.2 & 12006.2 \\ \hline
\end{tabular}
\end{center}
\caption{The contribution to the effective $\chi^2$ from each likelihood when fitting the EDE model to \baseline~and \baselinewoee~in combination with the \bao~and the \desi~datasets. Shorthand notations for the likelihoods are given in \cref{tab:likelihoods}.}
\label{tab:ede_bao_desi_chi2}
\end{table}

\begin{table}[]
\begin{center}
\begin{tabular} {| l | c| c|}
\hline
 \multicolumn{1}{|c|}{\bf } &  \multicolumn{1}{|c|}{\bf \baseline +\bao +\shoes} &  \multicolumn{1}{|c|}{\bf \baselinewoee +\bao +\shoes}\\
\hline
\PTT & 21.4 & 21.7 \\
 \PEE & 396.3 & \na \\ 
\PTTTEEE & 10549.2 & 10545.3 \\ 
 \act & 21.1 & 20.8 \\ 
 \sixdf & 0.00 & 0.02 \\
\mgs & 1.8 & 2.2 \\ 
\consensus & 3.5 & 3.5 \\ 
 \pantheonshoes & 1460.7 & 1460.7 \\ \hline
$\chi^2_{\mathrm{total}}$ & 12453.0 & 12054.2 \\ \hline
\end{tabular}
\end{center}
\caption{The contribution to the effective $\chi^2$ from each likelihood when fitting the EDE model to \baseline~and \baselinewoee~in combination with the \shoes~dataset. Shorthand notations for the likelihoods are given in \cref{tab:likelihoods}.}
\label{tab:ede_bao_shoes_chi2}
\end{table}

\begin{table}[]
\begin{center}
\begin{tabular} { |l | l|}
\hline
{\bf Likelihood} & {\bf Shorthand}\\
\hline
planck\_2018\_lowl.TT & \PTT\\
planck\_2018\_lowl.EE  &\PEE \\
planck\_NPIPE\_highl\_CamSpec.TTTEEE &\PTTTEEE  \\
act\_dr6\_lenslike.ACTDR6LensLike &\act \\
bao.sixdf\_2011\_bao &\sixdf\\
bao.sdss\_dr7\_mgs &\mgs\\
bao.sdss\_dr12\_consensus\_bao&\consensus\\
bao.desi\_2024\_bao\_all&\BAOdesi\\
sn.pantheonplus&\pantheon \\
sn.pantheonplusshoes& \pantheonshoes \\
 \hline
\end{tabular}
\end{center}
\caption{Shorthand notation used in the above tables of $\chi^2$ values to denote specific likelihoods as implemented in \texttt{Cobaya}.}
\label{tab:likelihoods}
\end{table}
    
\end{document}